\documentclass[12pt]{article}
\usepackage{amsmath}
\usepackage{amssymb}
\usepackage{amsthm}
\usepackage{graphicx}
\usepackage{lineno}
\usepackage{float}
\usepackage[all]{xy}
\usepackage{mathrsfs}

\usepackage{xcolor}
\usepackage{tikz}
\usetikzlibrary{tqft}

\newtheorem{theorem}{Theorem}
\newtheorem*{theorem*}{Theorem}

\newtheorem{definition}{Definition}

\newtheorem{example}{Example}

\newcommand{\tr}{{\rm tr}}

\newcommand{\A}{\mathcal{A}}
\newcommand{\B}{\mathcal B}
\newcommand{\Ce}{\mathcal C}

\newcommand{\lra}{\longrightarrow}

\newcommand{\medn}{\medbreak \noindent}
\newcommand{\bign}{\bigbreak \noindent}

\begin{document}

\title{\bf %Communication protocols and quantum error-correcting codes from the perspective of topological quantum field theory
Communication protocols and QECCs from the perspective of TQFT, Part I: \\
Constructing LOCC protocols and QECCs from TQFTs}

\author{{Chris Fields$^a$, James F. Glazebrook$^{b,c}$ and Antonino Marcian\`{o}$^{d,e,f}$}\\ \\
{\it$^a$ Allen Discovery Center, Tufts University, Medford, MA 02155 USA}\\
{fieldsres@gmail.com}\\
{ORCID: 0000-0002-4812-0744}\\
{\it$^b$ Department of Mathematics and Computer Science,} \\
{\it Eastern Illinois University, Charleston, IL 61920 USA} \\
{\it$^c$ Adjunct Faculty, Department of Mathematics,}\\
{\it University of Illinois at Urbana-Champaign, Urbana, IL 61801 USA}\\
{jfglazebrook@eiu.edu}\\
{ORCID: 0000-0001-8335-221X}\\
{\it$^d$ Center for Field Theory and Particle Physics \& Department of Physics} \\
{\it Fudan University, Shanghai, CHINA} \\
{marciano@fudan.edu.cn} \\
{\it$^e$ Laboratori Nazionali di Frascati INFN, Frascati (Rome), Italy, EU} \\
{marciano@lnf.infn.it} \\
{\it$^f$ INFN sezione Roma ``Tor Vergata'', 00133 Rome, Italy, EU} \\
{ORCID: 0000-0003-4719-110X}
}

\maketitle

{\bf Abstract:} \\

Topological quantum field theories (TQFTs) provide a general, minimal-assumption language for describing quantum-state preparation and measurement. They therefore provide a general language in which to express multi-agent communication protocols, e.g. local operations, classical communication (LOCC) protocols.  Here we construct LOCC protocols using TQFT, and show that LOCC protocols generically induce quantum error-correcting codes (QECCs). Using multi-observer scenarios described by quantum Darwinism and Bell/EPR experiments as examples, we show how these LOCC-induced QECCs effectively convert entanglement into classical redundancy.  In the accompanying Part II, we show that such QECCs can be regarded as implementing, or inducing the emergence of, spacetimes on the boundaries between interacting systems. We investigate this connection between inter-agent communication and spacetime using BF and Chern-Simons theories, and then using topological M-theory.

\tableofcontents

\section{Introduction}

We have previously shown \cite{fgm:22} that sequential observations of any finite physical system $S$ that employ either one or some sequence of quantum reference frames (QRFs) \cite{aharonov:84, bartlett:07} induce a topological quantum field theory (TQFT) \cite{atiyah:88, quinn:95} on $S$.  Briefly, we have first shown that the action of any QRF on any finite system $S$ can be represented by a particular category-theoretic construct of logically regulated information flow represented by a finite cone-cocone diagram (CCCD) \cite{fg:19a,fg:20, fgm:21} of Barwise-Seligman classifiers \cite{barwise:97}, which are effectively operators that assign ``tokens'' in some language to ``types'' in that language with defined probabilities.  These CCCDs are realizable as hierarchical `memory-read/write' distributed computations, functioning as logical gates, that implement each given QRF.  The basic ideas of this QRF to CCCD mapping are reviewed in Appendix A (see also \cite{fgm:22, fg:19a, fg:22}).  The second step taken in \cite{fgm:22} has been to construct a functor $\mathfrak{f}: \mathbf{CCCD} \rightarrow \mathbf{Cob}$ from the category $\mathbf{CCCD}$ of CCCDs representing QRFs to the category $\mathbf{Cob}$ of finite cobordisms.  Treating finite cobordisms as manifolds of Hilbert-space automorphisms in the usual way \cite{atiyah:88, quinn:95}, this functor maps morphisms of $\mathbf{CCCD}$ representing sequences of actions of QRFs on $S$ to a TQFT with copies of the Hilbert space $\mathcal{H}_S$ as its boundaries --- this construction is briefly reviewed in Appendix B.  A TQFT being the ``default'' physical theory induced on any system by measurement is not surprising: it reflects the fact that spacetime coordinates must, in any sequence of measurements, be specified by particular QRFs.  An immediate consequence of this construction is that any effective field theory (EFT) defined on $S$ must be gauge-invariant \cite{fgm:22, addazi:21}.

Here we investigate a further consequence of the existence of the functor $\mathfrak{f}: \mathbf{CCCD} \rightarrow \mathbf{Cob}$: any sequence of actions of, or equivalently, operations with, one or more QRFs can simply be identified with the TQFT that it induces.  That this should be the case is also not surprising.  A QRF is a physical system $Q$ with some internal dynamics representable by a Hamiltonian $H_Q$.  Hence treating it as isolated, it evolves in time $t$ via a unitary operator $\mathcal{P}_Q = \exp[(\imath / \hbar) H_Q t]$.  The ``boundaries'' between which $Q$ can be considered isolated are precisely its actions on $S$ at some sequential times $t_i, t_j, \dots t_k$.  Evolving $S$ through time and evolving $Q$ through time are, therefore, operationally the same process: they yield precisely the same data, the data obtained by acting with $Q$ on $S$ at $t_i, t_j, \dots t_k$.  This is in fact a bulk-boundary duality, with $Q$ the bulk and $\mathcal{H}_S$ the boundary \cite{fgm:22, fgm:22a}.

We show in what follows that the identification of QRFs with the TQFTs that they induce is theoretically powerful.  We use it in \S \ref{sharing} to construct a novel, purely topological proof that QRF sharing induces entanglement, a result previously demonstrated via the no-cloning theorem \cite{fgm:21, ffgl:22}.  We then show in \S \ref{protocols} that it provides a purely topological representation of multi-party communication protocols involving both quantum and classical resources, i.e. {\em Local Operations, Classical Communication} (LOCC) protocols \cite{chitambar:14}, and demonstrate that quantum Darwinism \cite{zurek:06, zurek:09}, originally formulated to explain the emergence of a ``public'' quantum-to-classical transition, simply describes a LOCC protocol. In \S \ref{qecc} we employ the generic, minimal-assumption formalism of Knill and Laflamme \cite{knill:97} to demonstrate our main result: that interactions meeting the requirements to implement a LOCC protocol generically induce quantum error correcting codes (QECCs).  We employ quantum Darwinism and Bell/EPR experiments as examples to investigate the effective conversion of entanglement into classical redundancy via a QECC.

In the accompanying Part II, we consider spacetime as a resource for classical redundancy, and show that QECCs generically induce spacetimes on the boundaries between interacting systems. We consider BF and Chern-Simons theories as examples, and show how spacetime considered as a QECC provides the redundancy required for LOCC.  Finally we show how these same considerations can be reformulated in the language of topological M-theory, and provide several specific examples.  We conclude that representing multi-party communication using TQFTs provides a new perspective on the connection between redundancy and spacetime, and suggests, consistent with remarks made already in \cite{addazi:21}, that redundancy and spacetime may at some fundamental level be the same concept.

\section{QRF sharing induces entanglement} \label{sharing}

As QRFs are by definition physical systems, their (quantum) states encode unmeasurable quantum phase information, referred to in \cite{bartlett:07} as ``nonfungible'' information.  The state --- in particular, the initial or ``ready'' state --- of a QRF cannot, therefore, be determined by any finite number of finite-resolution observations.  Whether, in particular, the prepared state of a QRF is identical to a theoretically-specified state cannot be determined by finite observations.  Unknown quantum states cannot, in general, be cloned by any unitary process \cite{wooters:82}.  In practice, therefore, the QRF sharing required by quantum communication protocols -- e.g. sharing an independently-measurable $z$ axis -- can only be approximate.  Here we develop an alternative proof of this result.

Let $U = AB$ be a bipartite decomposition of a closed system $U$ that satisfies the separability constraint $|AB \rangle = |A \rangle |B \rangle$ over any time period of interest; this can be achieved provided the interaction $H_{AB}$ is sufficiently weak.  Under these conditions, the decompositional boundary functions as a holographic screen $\mathscr{B}$.  This $\mathscr{B}$ is a discrete topological space comprising $N$ mutually-disjoint sites, where $2^N = \dim(H_{AB})$.  These $N$ sites can be considered to each house a single qubit, with the $N$ qubits collectively encoding the $2^N$ eigenvalues of $H_{AB}$ \cite{fgm:22, fg:19a, fg:20, fgm:21, addazi:21}.  The two systems $A$ and $B$ can be considered ``agents'' (Alice and Bob) that communicate by exchanging messages across $\mathscr{B}$; formally, we can consider them to alternately ``prepare'' and then ``measure'' the qubits on $\mathscr{B}$.  This situation is illustrated in Diagram \eqref{bipartite}.  Note that as the Hilbert space $\mathcal{H}_U = \mathcal{H}_A \otimes \mathcal{H}_B$, these $N$ qubits are ancillary to $\mathcal{H}_U$ and to the self-interaction $H_U$.

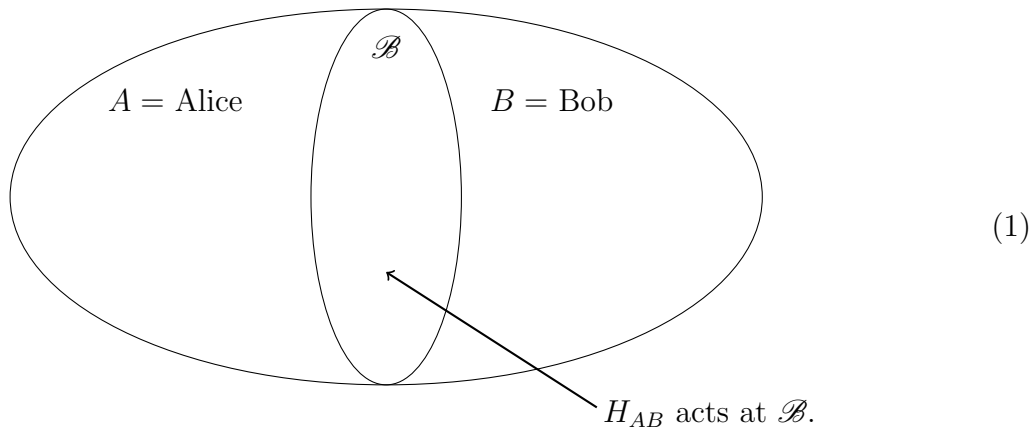
\begin{equation} \label{bipartite}
\begin{gathered}
\begin{tikzpicture}[every tqft/.append style={transform shape}]
\draw[rotate=90] (0,0) ellipse (2.5cm and 1 cm);
\node[above] at (0,1.7) {$\mathscr{B}$};
\draw (0,0) ellipse (5 cm and 2.5 cm);
\node[above] at (-2.8,1) {$A =$ Alice};
\node[above] at (2.2,1) {$B =$ Bob};
\node[above] at (4.3,-3.2) {$H_{AB}$ acts at $\mathscr{B}$.};
\draw [thick, ->] (2.8,-2.8) -- (0,-1);
\end{tikzpicture}
\end{gathered}
\end{equation}

Suppose now that $A$ and $B$ operate on $\mathscr{B}$ with QRFs $Q_m$ and $Q_m^\prime$, respectively; here we follow \cite{fgm:22} in using the notation `$Q$' to denote both a QRF and the dual preparation and measurement operations performed using that QRF. The domains ${\rm dom}(Q_m)$ and ${\rm dom}(Q_m^\prime)$ are subsets of qubits that encode the ``states'' of measured (dually, prepared) ``systems'' (or ``messages'') $S$ and $S^\prime$, respectively.  Suppose further that $A$ and $B$ record their observational outcomes on $\mathscr{B}$ with QRFs $Q_r$ and $Q_r^\prime$, respectively; the domains ${\rm dom}(Q_r)$ and ${\rm dom}(Q_r^\prime)$ are subsets of qubits that function as ``memories'' $Y$ and $Y^\prime$, respectively.  For convenience, we will assume $\dim(S) = \dim(Y)$ and $\dim(S^\prime) = \dim(Y^\prime)$.  The composite operation $Q = (\overrightarrow{Q}, \overleftarrow{Q})$, where $\overrightarrow{Q} = Q_r Q_m$ and $\overleftarrow{Q} = Q_m Q_r$, is then a pair of QRF sequences that can be identified with TQFTs that measure and record an outcome, mapping $\mathcal{H}_S \rightarrow \mathcal{H}_Y$, and dually use an outcome read from memory to prepare a state, mapping $\mathcal{H}_Y \rightarrow \mathcal{H}_S$, respectively; the composite operation $Q^\prime$ can be defined similarly.  These composite operations make explicit Wheeler's --- indeed Bohr's --- point that a quantum measurement cannot be considered to have occurred until it is irreversibly recorded \cite{wheeler:83}, in this case, written to the memory $Y$.  They can be represented as:

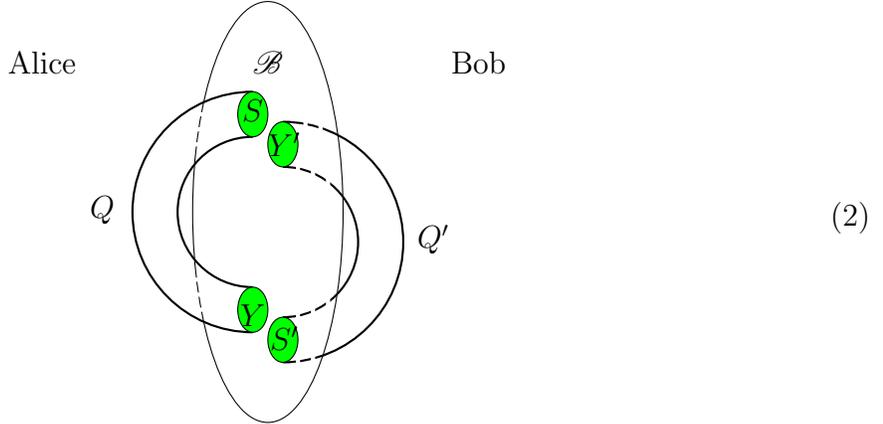
\begin{equation} \label{bipartite-ops}
\begin{gathered}
\begin{tikzpicture}[every tqft/.append style={transform shape}]
\draw[rotate=90] (0,0) ellipse (2.8cm and 1 cm);
\node[above] at (0,1.7) {$\mathscr{B}$};
\draw [thick] (-0.2,1.6) arc [radius=1.6, start angle=90, end angle= 270];
\draw [thick] (-0.2,1) arc [radius=1, start angle=90, end angle= 270];
\draw[rotate=90,fill=green,fill opacity=1] (1.3,0.2) ellipse (0.3 cm and 0.2 cm);
\draw[rotate=90,fill=green,fill opacity=1] (-1.3,0.2) ellipse (0.3 cm and 0.2 cm);
\draw [rotate=180, thick] (-0.2,2) arc [radius=1.6, start angle=90, end angle= 270];
\draw [rotate=180, thick] (-0.2,1.4) arc [radius=1, start angle=90, end angle= 270];
\draw[rotate=90,fill=green,fill opacity=1] (0.9,-0.2) ellipse (0.3 cm and 0.2 cm);
\draw[rotate=90,fill=green,fill opacity=1] (-1.7,-0.2) ellipse (0.3 cm and 0.2 cm);
\node[above] at (-3,1.7) {Alice};
\node[above] at (2.8,1.7) {Bob};
\node[above] at (-2.2,-0.3) {$Q$};
\node[above] at (2.2,-0.7) {$Q^\prime$};
\node at (-0.2,1.35) {$S$};
\node at (0.22,0.9) {$Y^{\prime}$};
\node at (-0.2,-1.37) {$Y$};
\node at (0.22,-1.68) {$S^{\prime}$};
\draw [ultra thick, white] (-0.8,1.3) -- (-1,1.3);
\draw [ultra thick, white] (-0.8,1.1) -- (-1,1.1);
\draw [ultra thick, white] (-0.8,0.9) -- (-1,0.9);
\draw [ultra thick, white] (-0.8,-0.9) -- (-1,-0.9);
\draw [ultra thick, white] (-0.8,-1.1) -- (-1,-1.1);
\draw [ultra thick, white] (-0.8,-1.3) -- (-1,-1.3);
\draw [ultra thick, white] (0.5,1.3) -- (0.5,1);
\draw [ultra thick, white] (0.7,1.3) -- (0.7,1);
\draw [ultra thick, white] (0.4,0.7) -- (0.4,0.4);
\draw [ultra thick, white] (0.6,0.6) -- (0.6,0.3);
\draw [ultra thick, white] (0.8,0.5) -- (0.8,0.2);
\draw [ultra thick, white] (0.6,-1.8) -- (0.6,-2.1);
\draw [ultra thick, white] (0.4,-1.9) -- (0.4,-2.1);
\draw [ultra thick, white] (0.4,-1.5) -- (0.4,-1.2);
\draw [ultra thick, white] (0.6,-1.4) -- (0.6,-1.1);
\draw [ultra thick, white] (0.8,-1.3) -- (0.8,-1);
\end{tikzpicture}
\end{gathered}
\end{equation}

where Alice and Bob are represented as the components of the bulk to the left and right of $\mathscr{B}$, respectively.  In this Diagram, Alice's and Bob's QRFs access different systems $S$ and $S^\prime$ and different memories $Y$ and $Y^\prime$, and can be considered completely independent.  They are, in particular, conditionally independent when $\mathscr{B}$ is replaced with a classical Markov blanket (MB), \cite{pearl:88} --- see \cite{ffgl:22} for details\footnote{Viewing $\mathscr{B}$  as a holographic screen, to see how it functions as an MB separating $A$ from $B$, we regard $\mathscr{B}$ as having an $2^N$-dimensional, $N$-qubit Hilbert space $\mathcal{H}_{q_i} = \prod_i q_i$.  While $\mathcal{H}_{q_i}$ is strictly ancillary to $\mathcal{H}_U = \mathcal{H}_A \otimes \mathcal{H}_B$, the classical situation can be recovered in the limit in which the entanglement entropies $\mathcal{S}(|A \rangle), \mathcal{S}(|B \rangle) \rightarrow 0$ by considering the products $\mathcal{H}_A \otimes \mathcal{H}_{q_i}$ and $\mathcal{H}_B \otimes \mathcal{H}_{q_i}$ to be ``particle'' state spaces for $A$ and $B$, respectively, for which, in this classical limit, the states of $\mathcal{H}_{q_i}$ then become the blanket states of an MB, which is trained inductively by conditional independence \cite{fgm:22,addazi:21}.}.

Now consider the case in which $S = Y^\prime$ and $Y = S^\prime$:

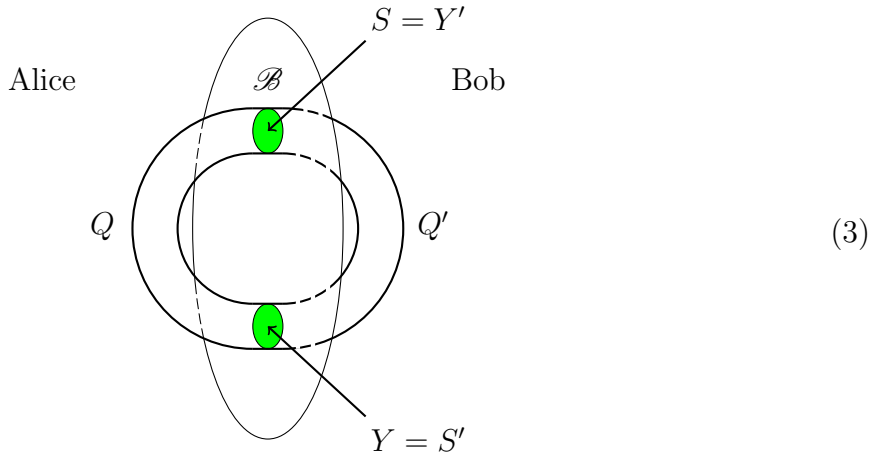
\begin{equation} \label{bipartite-share}
\begin{gathered}
\begin{tikzpicture}[every tqft/.append style={transform shape}]
\draw[rotate=90] (0,0) ellipse (2.8cm and 1 cm);
\node[above] at (0,1.7) {$\mathscr{B}$};
\draw [thick] (-0.2,1.6) arc [radius=1.6, start angle=90, end angle= 270];
\draw [thick] (-0.2,1) arc [radius=1, start angle=90, end angle= 270];
\draw[rotate=90,fill=green,fill opacity=1] (1.3,0) ellipse (0.3 cm and 0.2 cm);
\draw[rotate=90,fill=green,fill opacity=1] (-1.3,0) ellipse (0.3 cm and 0.2 cm);
\draw [rotate=180, thick] (-0.2,1.6) arc [radius=1.6, start angle=90, end angle= 270];
\draw [rotate=180, thick] (-0.2,1) arc [radius=1, start angle=90, end angle= 270];
\draw [thick] (-0.2,1.6) -- (0.2,1.6);
\draw [thick] (-0.2,1) -- (0.2,1);
\draw [thick] (-0.2,-1.6) -- (0.2,-1.6);
\draw [thick] (-0.2,-1) -- (0.2,-1);
\node[above] at (-3,1.7) {Alice};
\node[above] at (2.8,1.7) {Bob};
\node[above] at (-2.2,-0.3) {$Q$};
\node[above] at (2.2,-0.3) {$Q^\prime$};
\node at (2,2.8) {$S = Y^{\prime}$};
\draw [thick, ->] (1.3,2.5) -- (0,1.3);
\node at (2,-2.8) {$Y = S^{\prime}$};
\draw [thick, ->] (1.3,-2.5) -- (0,-1.3);
\draw [ultra thick, white] (-0.8,1.3) -- (-1,1.3);
\draw [ultra thick, white] (-0.8,1.1) -- (-1,1.1);
\draw [ultra thick, white] (-0.8,0.9) -- (-1,0.9);
\draw [ultra thick, white] (-0.8,-0.9) -- (-1,-0.9);
\draw [ultra thick, white] (-0.8,-1.1) -- (-1,-1.1);
\draw [ultra thick, white] (-0.8,-1.3) -- (-1,-1.3);
\draw [ultra thick, white] (0.5,1.7) -- (0.5,1.4);
\draw [ultra thick, white] (0.7,1.7) -- (0.7,1.4);
\draw [ultra thick, white] (0.4,1.1) -- (0.4,0.8);
\draw [ultra thick, white] (0.6,1) -- (0.6,0.7);
\draw [ultra thick, white] (0.8,0.9) -- (0.8,0.6);
\draw [ultra thick, white] (0.6,-1.4) -- (0.6,-1.7);
\draw [ultra thick, white] (0.4,-1.5) -- (0.4,-1.6);
\draw [ultra thick, white] (0.4,-1.1) -- (0.4,-0.8);
\draw [ultra thick, white] (0.6,-1) -- (0.6,-0.7);
\draw [ultra thick, white] (0.8,-0.9) -- (0.8,-0.6);
\end{tikzpicture}
\end{gathered}
\end{equation}

There are two consistent temporal flows in Diagram \eqref{bipartite-share}: $\overrightarrow{Q} \overrightarrow{Q^\prime}$ and $\overleftarrow{Q^\prime} \overleftarrow{Q}$, corresponding to each observer reading the other's memory and each observer measuring the other's system, respectively.  Both $\overrightarrow{Q} \overrightarrow{Q^\prime}$ and $\overleftarrow{Q^\prime} \overleftarrow{Q}$ are, however, TQFTs with copies of $S$ as boundaries.  By sharing QRF's, therefore, Alice and Bob jointly implement a single unitary process.  They are therefore entangled.

Diagram \eqref{bipartite-share} provides a new perspective on the no-cloning theorem, as it states that a unitary process cannot be ``cloned'' by separable observers.  When a process is viewed as an operator applied to a state, it becomes clear why this must be the case. As noted above, neither an unknown quantum state nor an unknown unitary operator can be fully specified by a finite set of finite-resolution observational outcomes.  While Alice and Bob may share {\em a priori} theoretical specifications of the states to be prepared, whether the prepared states meet these specifications exactly cannot be determined by finite observation.  All that Alice and Bob can learn about each other's state preparations is contained in the finite sets of observational outcomes that each can obtain by interacting with $\mathscr{B}$.  Specifically, they can obtain only the set of eigenvalues of $H_{AB}$ encoded at finite resolution on $\mathscr{B}$.  These observed eigenvalues are insufficient, in particular, to determine the entanglement entropy $\mathcal{S}(|AB \rangle )$ of the observers' joint state; see \cite{fg:23} Corollary 3.1 for proof.  Alice and Bob cannot, therefore, determine by observation whether they are entangled or mutually separable.

\section{Multi-party communication protocols} \label{protocols}

\subsection{Separability conditions on observers} \label{separability}

If separated observers cannot share QRFs, observational outcomes are observer-relative not as a matter of interpretation, but rather in principle \cite{fgm:22a}.  Hence quantum theory is naturally formulated as a single-observer theory \cite{bohr:34, everett:57, rovelli:96, mermin:98, tegmark:98, fuchs:10, mermin:18}.  In practice, however, we are mainly interested in situations in which two or more observers share an environment, including its information-processing resources, whether quantum or classical.  In such practical settings we require, moreover, the observers in question to be separable, i.e. that their joint state is separable at all times.

Let us suppose, then, that Alice is a bipartite system comprising two observers, $A_1$ and $A_2$, who interact with a shared environment, Bob.  We further suppose that $A_1$ and $A_2$ deploy composite QRFs $Q_1$ and $Q_2$.  Extending this decomposition to multiple observers is a simple, iterative process, but is hard on notation and graphically cumbersome; the important results can be obtained considering just two observers.

The QRFs $Q_1$ and $Q_2$ commute if and only if their representations as CCCDs commute, i.e. if and only if their representations as CCCDs have a mutual limit and a mutual colimit \cite{fgm:22}.  Indeed non-commutativity of QRFs induces quantum contextuality \cite{fgm:22} --- see \cite{fg:22} Thm. 3.4 for details.
\footnote{There are several interpretations of quantum contextuality which are putatively equivalent. For instance \cite{jaeger:20}: the dependency of measurement outcomes on circumstances external to the measured quantum system, or, for two system states $\mathbf{A}_1,\mathbf{A}_2$, the  joint measurement
$\mathbf{A}_1\mathbf{A}_2$ does not yield the same measurement statistics for $\mathbf{A}_1$ and $\mathbf{A}_2$ when these are measured separately
(for a comprehensive review of the concept, see \cite{adlam:21}).}

Here we show that if $Q_1$ and $Q_2$ commute, $A_1$ and $A_2$ are entangled, and so cannot be considered distinct observers.  This immediately implies that {\em all} observations recorded by distinct, i.e. mutually-separable observers are context-dependent.  We first demonstrate this result and show how it applies in a canonical case of joint manipulation of entanglement as a resource, then show how decoherence explains the observational elusiveness of the implied contextuality.

We first consider the case in which $Q_1$ and $Q_2$ have distinct pairs of domains $S_1$ and $Y_1$ and $S_2$ and $Y_2$, respectively.  In this case, nothing is changed by considering the disjoint unions $S_1 \sqcup Y_1$ and $S_2 \sqcup Y_2$ as the domains of $Q_1$ and $Q_2$, respectively.  Hence we have:

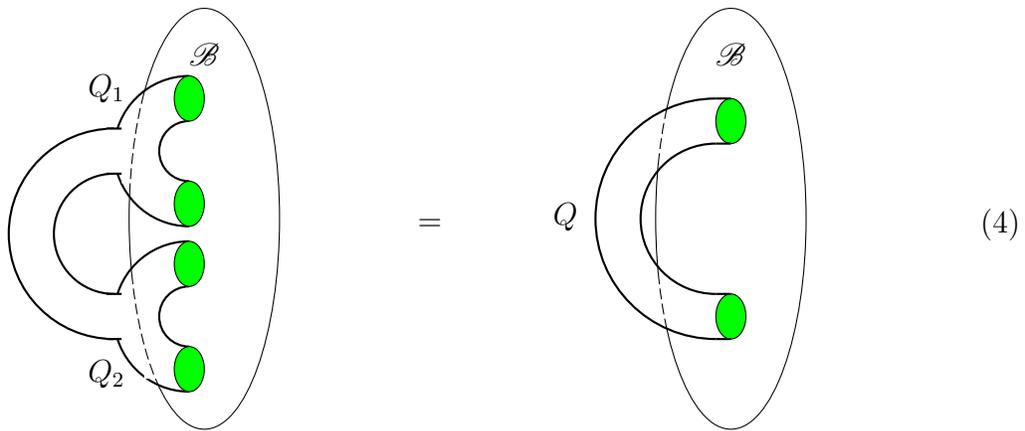
\begin{equation} \label{commute-QRF-1}
\begin{gathered}
\begin{tikzpicture}[every tqft/.append style={transform shape}]
\draw[rotate=90] (0,0) ellipse (2.8cm and 1 cm);
\node[above] at (0,1.9) {$\mathscr{B}$};
\draw [thick] (-0.2,1.6) arc [radius=1.6, start angle=90, end angle= 270];
\draw [thick] (-0.2,1) arc [radius=1, start angle=90, end angle= 270];
\draw[rotate=90,fill=green,fill opacity=1] (1.3,0) ellipse (0.3 cm and 0.2 cm);
\draw[rotate=90,fill=green,fill opacity=1] (-1.3,0) ellipse (0.3 cm and 0.2 cm);
\draw [thick] (-0.2,1.6) -- (0,1.6);
\draw [thick] (-0.2,1) -- (0,1);
\draw [thick] (-0.2,-1.6) -- (0,-1.6);
\draw [thick] (-0.2,-1) -- (0,-1);
\node[above] at (-2.2,-0.3) {$Q$};
\draw [ultra thick, white] (-0.8,1.3) -- (-1,1.3);
\draw [ultra thick, white] (-0.8,1.1) -- (-1,1.1);
\draw [ultra thick, white] (-0.8,0.9) -- (-1,0.9);
\draw [ultra thick, white] (-0.8,-0.9) -- (-1,-0.9);
\draw [ultra thick, white] (-0.8,-1.1) -- (-1,-1.1);
\draw [ultra thick, white] (-0.8,-1.3) -- (-1,-1.3);
\draw[rotate=90] (0,7) ellipse (2.8cm and 1 cm);
\node[above] at (-7,1.9) {$\mathscr{B}$};
\draw [thick] (-7.2,1.9) arc [radius=1, start angle=90, end angle= 270];
\draw [thick] (-7.2,1.3) arc [radius=0.4, start angle=90, end angle= 270];
\draw[rotate=90,fill=green,fill opacity=1] (1.6,7.2) ellipse (0.3 cm and 0.2 cm);
\draw[rotate=90,fill=green,fill opacity=1] (0.2,7.2) ellipse (0.3 cm and 0.2 cm);
\draw [thick] (-7.2,-0.3) arc [radius=1, start angle=90, end angle= 270];
\draw [thick] (-7.2,-0.9) arc [radius=0.4, start angle=90, end angle= 270];
\draw[rotate=90,fill=green,fill opacity=1] (-0.6,7.2) ellipse (0.3 cm and 0.2 cm);
\draw[rotate=90,fill=green,fill opacity=1] (-2.0,7.2) ellipse (0.3 cm and 0.2 cm);
\draw [thick] (-8.2,1.2) arc [radius=1.4, start angle=90, end angle= 270];
\draw [thick] (-8.2,0.6) arc [radius=0.8, start angle=90, end angle= 270];
\path [fill=white] (-8.3,1.2) rectangle (-8.1,0.6);
\path [fill=white] (-8.3,-1.6) rectangle (-8.1,-1);
\draw [thick] (-8.3,1.2) -- (-8.1,1.2);
\draw [thick] (-8.3,0.6) -- (-8.1,0.6);
\draw [thick] (-8.3,-1.6) -- (-8.1,-1.6);
\draw [thick] (-8.3,-1) -- (-8.1,-1);
\draw [ultra thick, white] (-7.9,1.5) -- (-7.7,1.5);
\draw [ultra thick, white] (-8,1.3) -- (-7.8,1.3);
\draw [ultra thick, white] (-8,1.1) -- (-7.8,1.1);
\draw [ultra thick, white] (-8,0.9) -- (-7.8,0.9);
\draw [ultra thick, white] (-8.1,0.7) -- (-7.8,0.7);
\draw [ultra thick, white] (-8.1,0.5) -- (-7.8,0.5);
\draw [ultra thick, white] (-8,-0.9) -- (-7.8,-0.9);
\draw [ultra thick, white] (-8,-1.1) -- (-7.8,-1.1);
\draw [ultra thick, white] (-8,-1.3) -- (-7.8,-1.3);
\draw [ultra thick, white] (-7.9,-1.5) -- (-7.7,-1.5);
\draw [ultra thick, white] (-7.9,-1.7) -- (-7.7,-1.7);
\draw [ultra thick, white] (-7.8,-1.9) -- (-7.6,-1.9);
\draw [ultra thick, white] (-7.8,-2.1) -- (-7.6,-2.1);
\node[above] at (-8.3,1.4) {$Q_1$};
\node[above] at (-8.3,-2.4) {$Q_2$};
\node[above] at (-4,-0.3) {=};
\end{tikzpicture}
\end{gathered}
\end{equation}

where we make no assumptions about QRFs deployed by Bob, but treat $H_B$ as implementing a generic quantum process.  The right-hand side of Diagram \eqref{commute-QRF-1} depicts a single TQFT, and hence a unitary process, implemented by Alice.  $A_1$ and $A_2$ are, therefore, entangled.  An explicit construction of this result employing the CCCD representation of the relevant QRFs and the algebraic operation of ``concurrence'' on CCCDs is provided in Appendix A.5.  This concurrence operation provides a general model of computational concurrency with a defined semantics, as summarized in Appendix A.2.

The second case of interest is that in which the domains of $Q_1$ and $Q_2$ are not independent, but rather overlap.  Suppose $S_1 = S_2 = S$; all other domain-overlap cases are similar.  Commutativity of $Q_1$ and $Q_2$ implies that $A_1$ can prepare $S$ simultaneously with $A_2$'s measurement of $S$.  Here again, $A_1$ and $A_2$ share a single unitary process and are, therefore, entangled.  As in the case of Diagram \eqref{bipartite-share}, $A_1$ and $A_2$ cannot measure their joint-state entanglement entropy; hence they cannot determine by observation that they are entangled, and hence cannot determine by observations that $Q_1$ and $Q_2$ commute.

Maintaining the distinction, i.e. separability, between $A_1$ and $A_2$ requires, therefore, that $Q_1$ and $Q_2$ do not commute.  In this case, $A_2$'s outcome probabilities for measurements of $S_2$ depend on $A_1$'s preparations of $S_1$ and vice-versa.  This is contextuality by default, not as an operational assumption \cite{dzha:17b} but as a first-principles requirement --- cf. \cite{abramsky:11} and \cite[\S7]{fg:22} \cite[\S7.2]{fgm:22}. Contextuality is, however, difficult to observe in practice, as reviewed in \cite{adlam:21,abramsky:18}.  Diagram \eqref{bipartite} makes it clear why: over the extended cycles of measurement required to accumulate statistics, Bob's internal dynamics $H_B$ mixes pure states $|S_1 \rangle$ and $|S_2 \rangle$ to decoherent densities $\rho_{S_1}$ and $\rho_{S_2}$.  $A_1$ and $A_2$ do not, in general, experience contextuality because they cannot, in general, measure a sufficiently large fraction of $\mathscr{B}$.

\subsection{LOCC protocols} \label{locc}

All communication protocols assume distinct, i.e. separable agents/observers capable of exchanging classical information via a shared language.  As a language ability can be taken to be implemented by a QRF \cite{fgm:21, ffgl:22}, the results of \S \ref{sharing} require that separable agents can at most approximately share a language.  Classical communication is enacted by encoding messages into and decoding messages from some physical medium, e.g. the ambient photon field.  Such media are quantum systems comprising degrees of freedom of the environment shared by the communicating agents (cf. the discussion in \cite{tipler:14}); in the present notation, they are components of Bob.  A medium implementing an optimal, noise-free classical channel can be represented as a QRF implemented by Bob.  Such a channel introduces, at most, a basis rotation of the encoded data; such a basis rotation is sufficient to assure non-commutativity of the sending and receiving QRFs.  Treating the message space as a shared classical memory, we can represent this channel QRF as a TQFT mapping $Y_1 \leftrightarrows Y_2$.  Hence we can represent classical communication by:

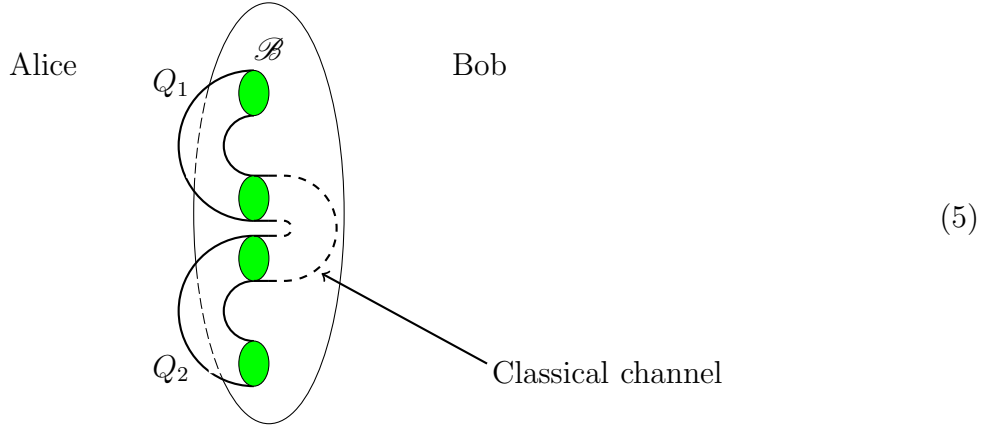
\begin{equation} \label{classical}
\begin{gathered}
\begin{tikzpicture}[every tqft/.append style={transform shape}]
\draw[rotate=90] (0,0) ellipse (2.8cm and 1 cm);
\node[above] at (0,1.9) {$\mathscr{B}$};
\draw [thick] (-0.2,1.9) arc [radius=1, start angle=90, end angle= 270];
\draw [thick] (-0.2,1.3) arc [radius=0.4, start angle=90, end angle= 270];
\draw[rotate=90,fill=green,fill opacity=1] (1.6,0.2) ellipse (0.3 cm and 0.2 cm);
\draw[rotate=90,fill=green,fill opacity=1] (0.2,0.2) ellipse (0.3 cm and 0.2 cm);
\draw [thick] (-0.2,-0.3) arc [radius=1, start angle=90, end angle= 270];
\draw [thick] (-0.2,-0.9) arc [radius=0.4, start angle=90, end angle= 270];
\draw[rotate=90,fill=green,fill opacity=1] (-0.6,0.2) ellipse (0.3 cm and 0.2 cm);
\draw[rotate=90,fill=green,fill opacity=1] (-2.0,0.2) ellipse (0.3 cm and 0.2 cm);
\draw [rotate=180, thick, dashed] (-0.2,0.9) arc [radius=0.7, start angle=90, end angle= 270];
\draw [rotate=180, thick, dashed] (-0.2,0.3) arc [radius=0.1, start angle=90, end angle= 270];
\draw [thick] (-0.2,0.5) -- (0,0.5);
\draw [thick] (-0.2,-0.1) -- (0,-0.1);
\draw [thick] (-0.2,-0.9) -- (0,-0.9);
\draw [thick] (-0.2,-0.3) -- (0,-0.3);
\draw [thick, dashed] (0,0.5) -- (0.2,0.5);
\draw [thick, dashed] (0,-0.1) -- (0.2,-0.1);
\draw [thick, dashed] (0,-0.9) -- (0.2,-0.9);
\draw [thick, dashed] (0,-0.3) -- (0.2,-0.3);
\node[above] at (-3,1.7) {Alice};
\node[above] at (2.8,1.7) {Bob};
\draw [ultra thick, white] (-0.9,1.5) -- (-0.7,1.5);
\draw [ultra thick, white] (-1,1.3) -- (-0.8,1.3);
\draw [ultra thick, white] (-1,1.1) -- (-0.8,1.1);
\draw [ultra thick, white] (-1,0.9) -- (-0.8,0.9);
\draw [ultra thick, white] (-1.1,0.7) -- (-0.8,0.7);
\draw [ultra thick, white] (-1.1,0.5) -- (-0.8,0.5);
\draw [ultra thick, white] (-1,-0.9) -- (-0.8,-0.9);
\draw [ultra thick, white] (-1,-1.1) -- (-0.8,-1.1);
\draw [ultra thick, white] (-1,-1.3) -- (-0.8,-1.3);
\draw [ultra thick, white] (-0.9,-1.5) -- (-0.7,-1.5);
\draw [ultra thick, white] (-0.9,-1.7) -- (-0.7,-1.7);
\draw [ultra thick, white] (-0.8,-1.9) -- (-0.6,-1.9);
\draw [ultra thick, white] (-0.8,-2.1) -- (-0.6,-2.1);
\node[above] at (-1.3,1.4) {$Q_1$};
\node[above] at (-1.3,-2.4) {$Q_2$};
\node[above] at (4.5,-2.4) {Classical channel};
\draw [thick, ->] (2.9,-2) -- (0.7,-0.8);
\end{tikzpicture}
\end{gathered}
\end{equation}

where $A_1$ and $A_2$ are assumed to be distinct agents and hence $Q_1$ and $Q_2$ are assumed not to commute.  Note that from an operational perspective, this assumption of non-commutativity corresponds to the non-observability of the channel QRF implemented by Bob, which is separated from $A_1$ and $A_2$ by $\mathscr{B}$.  $A_1$ and $A_2$ cannot, therefore, deduce the preparation conditions that this channel imposes on $Y_2$, or in the reverse direction, on $Y_1$.  This is in fact the case in ordinary classical communication between humans, and is why classical eavesdroppers can avoid detection.  From a classical perspective, the boundary $\mathscr{B}$ implements an MB separating the observers from the channel and restricting their access to it \cite{fgm:22, ffgl:22}.

In a LOCC protocol, $A_1$ and $A_2$ are assumed to know the preparation conditions for their shared quantum channel; indeed these preparation conditions are what they classically communicate.  Hence we can represent a LOCC protocol by adding a quantum channel -- a TQFT -- to Diagram \eqref{classical}:

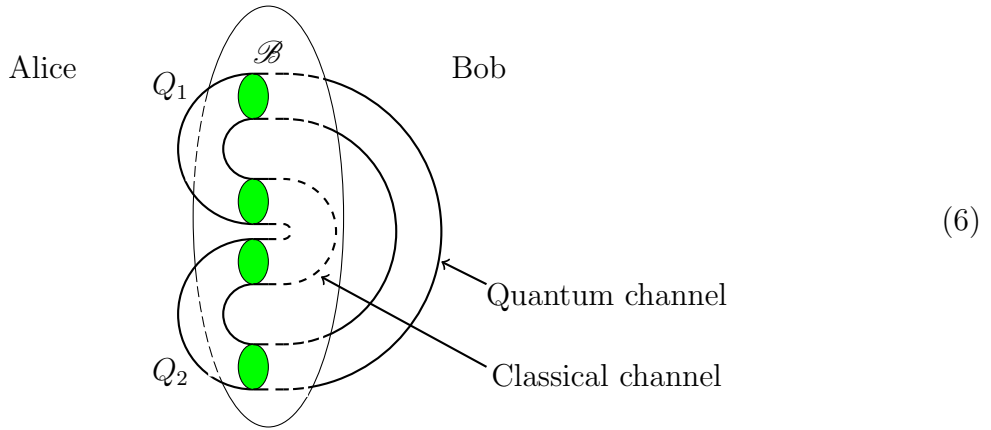
\begin{equation} \label{locc-diag}
\begin{gathered}
\begin{tikzpicture}[every tqft/.append style={transform shape}]
\draw[rotate=90] (0,0) ellipse (2.8cm and 1 cm);
\node[above] at (0,1.9) {$\mathscr{B}$};
\draw [thick] (-0.2,1.9) arc [radius=1, start angle=90, end angle= 270];
\draw [thick] (-0.2,1.3) arc [radius=0.4, start angle=90, end angle= 270];
\draw[rotate=90,fill=green,fill opacity=1] (1.6,0.2) ellipse (0.3 cm and 0.2 cm);
\draw[rotate=90,fill=green,fill opacity=1] (0.2,0.2) ellipse (0.3 cm and 0.2 cm);
\draw [thick] (-0.2,-0.3) arc [radius=1, start angle=90, end angle= 270];
\draw [thick] (-0.2,-0.9) arc [radius=0.4, start angle=90, end angle= 270];
\draw[rotate=90,fill=green,fill opacity=1] (-0.6,0.2) ellipse (0.3 cm and 0.2 cm);
\draw[rotate=90,fill=green,fill opacity=1] (-2.0,0.2) ellipse (0.3 cm and 0.2 cm);
\draw [rotate=180, thick, dashed] (-0.2,0.9) arc [radius=0.7, start angle=90, end angle= 270];
\draw [rotate=180, thick, dashed] (-0.2,0.3) arc [radius=0.1, start angle=90, end angle= 270];
\draw [thick] (-0.2,0.5) -- (0,0.5);
\draw [thick] (-0.2,-0.1) -- (0,-0.1);
\draw [thick] (-0.2,-0.9) -- (0,-0.9);
\draw [thick] (-0.2,-0.3) -- (0,-0.3);
\draw [thick, dashed] (0,0.5) -- (0.2,0.5);
\draw [thick, dashed] (0,-0.1) -- (0.2,-0.1);
\draw [thick, dashed] (0,-0.9) -- (0.2,-0.9);
\draw [thick, dashed] (0,-0.3) -- (0.2,-0.3);
\node[above] at (-3,1.7) {Alice};
\node[above] at (2.8,1.7) {Bob};
\draw [ultra thick, white] (-0.9,1.5) -- (-0.7,1.5);
\draw [ultra thick, white] (-1,1.3) -- (-0.8,1.3);
\draw [ultra thick, white] (-1,1.1) -- (-0.8,1.1);
\draw [ultra thick, white] (-1,0.9) -- (-0.8,0.9);
\draw [ultra thick, white] (-1.1,0.7) -- (-0.8,0.7);
\draw [ultra thick, white] (-1.1,0.5) -- (-0.8,0.5);
\draw [ultra thick, white] (-1,-0.9) -- (-0.8,-0.9);
\draw [ultra thick, white] (-1,-1.1) -- (-0.8,-1.1);
\draw [ultra thick, white] (-1,-1.3) -- (-0.8,-1.3);
\draw [ultra thick, white] (-0.9,-1.5) -- (-0.7,-1.5);
\draw [ultra thick, white] (-0.9,-1.7) -- (-0.7,-1.7);
\draw [ultra thick, white] (-0.8,-1.9) -- (-0.6,-1.9);
\draw [ultra thick, white] (-0.8,-2.1) -- (-0.6,-2.1);
\node[above] at (-1.3,1.4) {$Q_1$};
\node[above] at (-1.3,-2.4) {$Q_2$};
\draw [rotate=180, thick] (-0.2,2.3) arc [radius=2.1, start angle=90, end angle= 270];
\draw [rotate=180, thick] (-0.2,1.7) arc [radius=1.5, start angle=90, end angle= 270];
\draw [thick] (-0.2,1.9) -- (0,1.9);
\draw [thick] (-0.2,1.3) -- (0,1.3);
\draw [thick, dashed] (0.2,1.9) -- (0,1.9);
\draw [thick, dashed] (0.2,1.3) -- (0,1.3);
\draw [thick] (-0.2,-1.7) -- (0,-1.7);
\draw [thick] (-0.2,-2.3) -- (0,-2.3);
\draw [thick, dashed] (0.2,-1.7) -- (0,-1.7);
\draw [thick, dashed] (0.2,-2.3) -- (0,-2.3);
\draw [ultra thick, white] (0.3,2) -- (0.3,1.2);
\draw [ultra thick, white] (0.5,2) -- (0.5,1.2);
\draw [ultra thick, white] (0.7,1.9) -- (0.7,1.1);
\draw [ultra thick, white] (0.3,-2.4) -- (0.3,-1.5);
\draw [ultra thick, white] (0.5,-2.4) -- (0.5,-1.5);
\draw [ultra thick, white] (0.7,-1.8) -- (0.7,-1.5);
\node[above] at (4.5,-2.4) {Classical channel};
\draw [thick, ->] (2.9,-2) -- (0.7,-0.8);
\node[above] at (4.5,-1.4) {Quantum channel};
\draw [thick, ->] (2.9,-0.9) -- (2.3,-0.6);
\end{tikzpicture}
\end{gathered}
\end{equation}

Note that the two channel QRFs implemented by Bob in Diagram \eqref{locc-diag} are functionally equivalent: the labels ``classical'' and ``quantum'' can be exchanged without altering the diagram.  This symmetry reveals the dependence of LOCC protocols on the assumption that communication between observers in language can be regarded as classical.  This assumption of classicality operationalizes the assumption of separability between observers that is required for the notions of ``joint observation'' and ``joint manipulation'' to be physically meaningful. Note also that the intersections of the classical information channels with $\mathscr{B}$ in \eqref{classical} and \eqref{locc-diag}, comprise the (classical) MBs of Alice and Bob, respectively.

\begin{example}{\textbf{Entanglement of formation as a measure of non-commutativity}}

{\rm Protocols that involve the manipulation of multiple entangled states provide a more elaborate example of LOCC that single-qubit exchange.  Wootters \cite{wootters:01} considers the case in which two observers jointly prepare a shared mixed state $\rho$ using shared singlet states as a resource.  Classical communication is clearly required; it is assumed that no additional quantum communication channel is available.  Adopting the notation of \cite{wootters:01} in which some set $\{ \Psi_i \}$ of singlet states are shared by parties $A$ and $B$, the number of shared singlets that must be consumed to create $n$ copies of some given pure state $|\Phi \rangle$ is $nE(\Phi)$, where $E(\Phi)$ is the von Neumann entanglement entropy $\mathcal{S}$ across the $A$ - $B$ partition \cite[Eq. 3]{wootters:01}:}

\begin{equation}\label{share-1}
E(\Phi) = \mathcal{S} (\tr_B \vert \Phi \rangle \langle \Phi |) = \mathcal{S} (\tr_A \vert \Phi \rangle \langle \Phi |) = - \sum^{n}_{i} c_i^2 \log_2c_i^2\,,
\end{equation}
\noindent
{\rm where the $c_i$ are Schmidt coefficients.   Suppose now that the goal is to create $n$ copies of a mixed state $\rho$ that can be decomposed into pure states as \cite[Eq. 4]{wootters:01}:}

\begin{equation}\label{share-2}
\rho = \sum^N_{j=1} p_j \vert \Phi_j \rangle \langle \Phi_j \vert\,.
\end{equation}
\noindent
{\rm Here the $\vert \Phi_j \rangle$ are distinct, but not necessarily orthogonal, normalized pure states of the bipartite ($A$, $B$) system, and $p_j \geq 0$ (with $\sum_j p_j = 1$). The number of singlets expended is \cite[Eq. 5]{wootters:01}:
\begin{equation}\label{share-2a}
n \sum^N_{j=1} p_j E(\Phi_j)
\end{equation}
with $E(\Phi_j)$ given by Eq. \eqref{share-1}.

The {\em entanglement of formation} of the mixed state $\rho$ is then defined by \cite[Eq. 6]{wootters:01}:
\begin{equation}\label{share-3}
E_f(\rho) = \inf \sum_j p_j E(\Phi_j)\,,
\end{equation}
with the infimum taken over all pure state decompositions of $\rho$.  Comparing Eq. \eqref{share-3} with \eqref{share-2a}, $E_f(\rho)$ is just the minimum number of singlets needed to make each copy of $\rho$.}

{\rm As pointed out in \cite{wootters:01}, the entanglement of formation is not recoverable by distillation; the conversion of $\{ \Psi_i \}$ to $\rho$ is an irreversible process.  The term $E_f(\rho)$ is, therefore, effectively a measure of the quantum information that this conversion renders inaccessible.  To see how this information is ``lost'' during the conversion, it is useful to consider the process as a LOCC protocol on the model of Diagram \eqref{locc-diag}, with $A_1$ and $A_2$ the agents performing the protocol and the state transition $\{ \Psi_i \} \rightarrow \rho$ being implemented by Bob in response to their manipulations.  In this setting, irreversibility corresponds to non-commutativity between $Q_1$ and $Q_2$ when interpreted as operators as discussed above.  We can, therefore, write:}

\begin{equation}
E_f(\rho) = (2/h) || [Q_1, Q_2] \rho || \,,
\end{equation}

{\rm with $h$ Planck's constant.  The upper limit $E_f (\Psi ) = E(\Psi ) = 1$ for a singlet state corresponds to the action of fixing one unknown bit, the one not measured by the first-acting operator.  The lower limit $E_f(\rho) = 0$ is achieved if and only if $\rho$ is separable \cite{wootters:01}, i.e. only if $A_1$ and $A_2$ do not share a quantum channel traversing Bob.  This is consistent with $A_1$ and $A_2$ remaining distinct agents only if the global decomposition $U = AB$ is replaced by $U = ZW$, where $Z = \mathrm{Alice}_1 \rho_1$, $W = \mathrm{Alice}_2 \rho_2$, and $\rho = \rho_1 \rho_2$, exemplifying the general decomposition dependence of entanglement and separability \cite{zanardi:01, zanardi:04, torre:10, harshman:11, thirring:11}.}
\end{example}

\begin{example}{\textbf{Undecidability of QRF sharing}}

{\rm A LOCC protocol can be viewed as a method that allows $A_1$ and $A_2$ to determine, via observation and classical communication, whether they share a quantum channel; it is this perspective on LOCC that is implemented by Bell/EPR experiments \cite{aspect:81, georgescu:21}.  It can also be viewed, via the symmetry of Diagram \eqref{locc-diag}, as allowing $A_1$ and $A_2$ to determine, via manipulations of a shared quantum resource, whether they share a classical communication channel, and hence a classical language.  A ``language'' is a shared mapping from input data to output data, which must be physically implemented by a QRF and tracked for semantic consistency in the accompanying information flow by a CCCD.  Hence we can view a LOCC protocol as allowing $A_1$ and $A_2$ to determine, via manipulations of a shared quantum resource, whether they share a QRF.  As we have seen, QRF sharing induces entanglement.  It is for this reason that, as noted above, $A_1$ and $A_2$ can assume that they share a common language for practical purposes, but they cannot demonstrate it by making local measurements.  $A_1$ cannot, in particular, deduce from local measurements the  QRF employed by $A_2$.  Nor can Bob deduce from measurements that $A_1$ and $A_2$ share a QRF.}

{\rm Restating this result in the language of formal undecidability reveals the deep connection between LOCC protocols and computability.  Besides the interpretation in terms of measurement/preparation operations given in \S\ref{sharing}, a QRF can also be thought as an operator $Q: \psi \mapsto P_{\psi} (x_i)$ that acts on a quantum state $\psi$ to yield a probability distribution $P_{\psi} (x_i)$ of QRF-relative outcome values $x_i$ with well-defined units of measurement.  If $Q$ is a QRF and $\psi$ and $\phi$ are quantum states, then for any unitary $U: \psi \mapsto \phi$, there must be a map $f_U: P_{\psi} (x_i) \mapsto P_{\phi} (x_i)$ such that the diagram: }

\begin{equation}\label{diag1}
\begin{gathered}
\xymatrix{
\psi~ \ar[d]_{Q} \ar@{-->}[r]^{U} & \phi \ar[d]^{Q}\\
P_{\psi} (x_i)  \ar[r]^{f_U} & P_{\phi} (x_i)
}
\end{gathered}
\end{equation}
~ \\
\noindent
{\rm commutes, i.e. $f_U Q = Q U$.  Any such $Q$ clearly respects identities and composition, i.e. if $U = VW$ there must be maps $f_U = f_V f_W$, so $Q$ can be considered functorial.  In practice, we are only interested in finite apparatus interacting with finite systems, hence we can write: }

\begin{equation*}
Q: \mathbf{QuantState}_N \rightarrow \mathbf{ProbDist}_N
\end{equation*}

\noindent
{\rm with $\mathbf{QuantState}_N$ the category of quantum states with some finite dimension $N$ (i.e. rays in the Hilbert space of dimension $N$) and $\mathbf{ProbDist}_N$ the category of probability distributions over a set of $N$ elements.  In this functorial representation, the nonfungibility of the physical system $Q$ corresponds to the forgetfulness of the functor $Q$: the phase information that determines, via $Q$, the function $f_U$ cannot be reconstructed from $f_U$ itself.  Indeed we can always find quantum states $\zeta$ and $\xi$, a unitary $V: \zeta \mapsto \xi$, and an alternative QRF $R \neq Q$ such that the diagram: }

\begin{equation}\label{diag2}
\begin{gathered}
\xymatrix{
\psi~ \ar[d]_{Q} \ar@{-->}[r]^{U} & \phi \ar[d]^{Q}\\
P_{\psi, \zeta} (x_i)  \ar[r]^{f_U}_{f_V} & P_{\phi, \xi} (x_i)\\
\zeta~ \ar[u]^{R} \ar@{-->}[r]_{V} & \xi \ar[u]_{R}
}
\end{gathered}
\end{equation}
~ \\
\noindent
{\rm commutes; here the notation $P_{a,b} (x_i)$ abbreviates $P_a (x_i) = P_b (x_i)$.  From a quantum computing perspective, this is trivial: it simply says that any function $f_U$ can be implemented to arbitrary accuracy by multiple quantum algorithms.}

{\rm Let us now assume that Bob has the computational capability of a Turing machine \cite{turing:37, hopcroft:79} with an arbitrarily large memory, and ask whether Bob can decide, given finite measurements of the states $\psi$, $\phi$, $\zeta$, and $\xi$ and finite descriptions of $Q$ and $R$ --- in the notation of Diagram \eqref{locc-diag}, $Q_1$ and $Q_2$ --- whether Diagram \eqref{diag2} commutes.  To make the question precise, we can assume that $Q$ and $R$ are described by finite programs in some Turing-complete programming language; the question is then whether Bob can decide whether $f_U = f_V$ given these two programs and a finite set of inputs.  Rice's theorem \cite{rice:53} answers this question in the negative: no Turing machine can decide what function is computed by an arbitrarily-given program (see Appendix C for discussion and proof).  By the same reasoning, neither $A_1$ nor $A_2$ can decide whether their respective QRFs compute the same function.  }

{\rm The separability between $A_1$ and $A_2$ on which LOCC protocols depend rests, therefore, on the undecidability of QRF sharing, and hence of entanglement.  One of the simplest LOCC protocols is replicate measurement; Rice's theorem implies that whether $A_2$ has replicated $A_1$'s measurement is undecidable.  In practice, confirming replication becomes increasingly difficult as the measurement resolution $N \rightarrow \infty$, i.e. as the functions $f_U$ and $f_V$ are specified with greater and greater precision.  Replication depends, in practice, on the coarse-graining imposed by macroscopic calibration procedures, effectively-classical initial states, and finite measurement resolution.  It depends, in other words, on replication being ``up to'' some classical noise.  This point was, of course, already made by Bohr in 1928 \cite{bohr:28}.        }
\end{example}

\subsection{Quantum Darwinism describes a LOCC protocol} \label{qDar}

Classical communication is an instance of environmental redundancy: distinct observers $A_1$ and $A_2$ can share a classical message only if the message is redundantly encoded on $\mathscr{B}$ by the action of the ``environmental'' dynamics $H_B$.  Redundant encoding of the quantum states of ``systems'' embedded in a ``witnessing'' environment \cite{zurek:03, zurek:05} is the basis of quantum Darwinism \cite{zurek:06, zurek:09}; see \cite{korbicz:21} for a recent review.  For any system $X$ embedded in the environment, the pointer states selected for encoding are the eigenstates of $H_{XE_{X}}$, where $E_X$ is the environment of $X$, i.e. $XE_X = B$.  When multiple observers $A_i$ are present, each observer exclusively accesses a component $E_{X_i}$ of $E_X$. The state $|X \rangle$ is encoded {\em redundantly} if each of the $A_i$, interacting exclusively with the single environmental component $E_{X_i}$, obtains the same measurement outcome, and hence reports an observation of the same state $|X \rangle$.  Note that while the encoding of $|X \rangle$ in $E_X$ is a quantum process, the observational reports that reveal the redundant encoding are classical.  Redundant encoding of $|X \rangle$ in multiple components $E_{X_i}$ of $E_X$ clearly requires that $H_{XE_{X_i}} = H_{XE_{X_j}}$ for all components $i, j$.  As the observers are assumed to act completely independently, they must be mutually separable; separability is preserved -- hence contextuality is prevented -- by requiring that the environmental sectors $E_{X_{i}}$ are disjoint and do not interact. Showing just two observers explicitly, the encoding process can be depicted as (cf. \cite{zurek:05} Fig. 1):

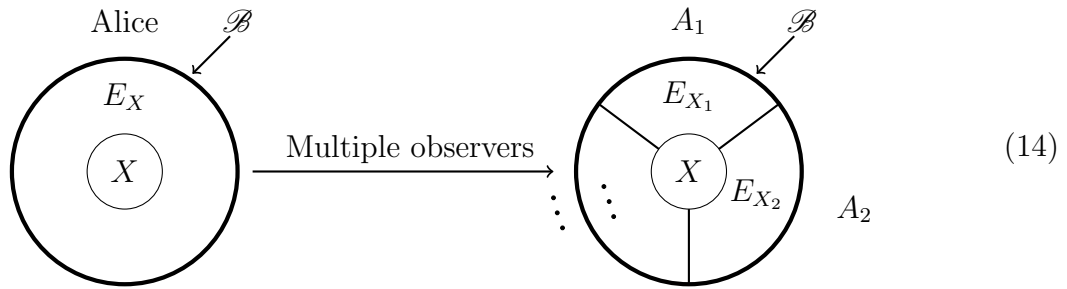
\begin{equation} \label{qdar}
\begin{gathered}
\begin{tikzpicture}
\draw (0,0) circle [radius = 0.5];
\draw [ultra thick] (0,0) circle [radius = 1.5];
\node at (0,0) {$X$};
\node at (0,1) {$E_X$};
\node at (0,2) {Alice};
\node at (1.5,2) {$\mathscr{B}$};
\draw [thick, ->] (1.4,1.8) -- (0.9,1.3);
\node at (3.8,0.3) {Multiple observers};
\draw [thick, ->] (1.7,0) -- (5.7,0);
\draw (7.5,0) circle [radius = 0.5];
\draw [ultra thick] (7.5,0) circle [radius = 1.5];
\node at (7.5,0) {$X$};
\node at (7.5,1) {$E_{X_{1}}$};
\node at (7.5,2) {$A_1$};
\node at (9,2) {$\mathscr{B}$};
\draw [thick, ->] (8.9,1.8) -- (8.4,1.3);
\draw [thick] (6.3,0.9) -- (7.1,0.3);
\draw [thick] (8.7,0.9) -- (7.9,0.3);
\draw [thick] (7.5,-0.5) -- (7.5,-1.5);
\node at (8.4,-0.3) {$E_{X_{2}}$};
\node at (9.7,-0.5) {$A_2$};
\draw[fill] (6.5,-0.6) circle [radius=0.025];
\draw[fill] (6.44,-0.4) circle [radius=0.025];
\draw[fill] (6.37,-0.2) circle [radius=0.025];
\draw[fill] (5.82,-0.75) circle [radius=0.025];
\draw[fill] (5.75,-0.55) circle [radius=0.025];
\draw[fill] (5.7,-0.35) circle [radius=0.025];
\end{tikzpicture}
\end{gathered}
\end{equation}

As formulated in \cite{zurek:06, zurek:09}, quantum Darwinism assumes that the observers can distinguish, and interact specifically with, the $X$-specific environmental fragments $E_{X_{i}}$ without prior communication or prior knowledge of $X$.  This assumption is clearly unrealistic \cite{fields:10}; the observers require $X$-specific QRFs to distinguish the $E_{X_{i}}$ from alternative environmental fragments that encode pointer states of other systems.  If we add this requirement, then each of the $E_{X_{i}}$ implements a quantum channel from $X$ to the relevant observer $A_i$, and hence composite systems of the form $E_{X_{i}} X E_{X_{j}}$ implement quantum channels between observers $A_i$ and $A_j$. If we further allow the observers to agree to observe $X$ and to share their observations via a separate, classical channel, we have a LOCC protocol.

The Quantum Darwinist construction highlights the critical role of redundant encoding of observationally-accessible data on $\mathscr{B}$ for systems $A$ comprising multiple observers.  If the observers are to remain distinct, and therefore mutually separable, the results of \S \ref{locc} require that the redundancy of encoding be only approximate, a requirement that is met if the different environmental fractions $E_{X_{i}}$ impose even slight basis rotations on the data they transport from $X$.  As amply demonstrated in practice, such approximately redundant encodings allow communities of observers, each interacting with the world via their own MB, i.e. their own sector of $\mathscr{B}$ in the classical limit, to correct observational errors.  Hence quantum Darwinism is, effectively, a mechanism for constructing an error-correcting code; we will see below that it is if fact a QECC.

%%%%%%%%%%%%%%%%%%%%%%%%%%%%%%%%%%%%%%%%%%%%%%%%%%%%%%%%%%%%%%%%%%%%%%%%%%%%%

\section{LOCC protocols induce QECCs} \label{qecc}

With this background, it is straightforward to see that any LOCC protocol induces a QECC, or equivalently, that any LOCC protocol executed by Alice requires that Bob implements a QECC.  Using the notation of Diagram \eqref{locc-diag}, $A_1$ and $A_2$ can communicate via a LOCC protocol only if Bob's internal interaction $H_B$ implements both quantum and classical channels with sufficient fidelity.  We can assume that the classical channel is implemented with sufficient fidelity to allow classical error-correction methods, e.g. parity checking or message repetition; either is just a matter of sufficient bandwidth.  The relevant question is assuring the fidelity of the quantum channel. The conditions for doing so are given under generic assumptions by the general theory of QECCs \cite{knill:97}, and involve encoding states to be protected into a larger Hilbert space subject to known interactions.  These conditions constitute, by definition, those that must generically be met by a QECC. Hence LOCC protocols require QECCs.

As discussed in \cite{knill:97}, the task of a QECC is to protect an encoded quantum state from degradation by the environment.  In the simplest case, we can regard the encoded quantum state as the state $|S \rangle$ of some sector $S$ of $\mathscr{B}$; in the limit, this state $|S \rangle$ may be the state of a single qubit, as in, e.g. one arm of a Bell/EPR experiment.  Treating $S$ as defined by, and hence $|S \rangle$ as prepared by, the ``user'' of the QECC, Alice, the relevant environment is the entirety of Bob, other than those components that implement the quantum channel and hence the QECC.  Adapting the notation of \cite{knill:97} to that of Diagram \eqref{locc-diag}, the perturbative action of the environment (i.e. Bob) can be represented by a set of error-inducing operators $B_{\alpha}$ such that \cite[Eq. 5]{knill:97}:
\begin{equation} \label{B-op-def}
\sum_{\alpha} B^{\dagger}_{\alpha} B_{\alpha} = I.
\end{equation}
\noindent
Letting $2^k$ be the dimension of the effective Hilbert space $\mathcal{H}_S$ of the sector $S$ to be preserved, the codespace is a $2^n$ dimensional Hilbert space $\mathcal{H}_{\mathcal{C}}$, where $n >> k$ and $\mathcal{C}$ designates the code.  An encoding operator is then a positive, trace-preserving, surjective map $E: S \rightarrow \mathcal{C}$; a decoding operator is its right inverse $D$, i.e. $ED = I$.  Letting $|0_L \rangle$ and $|1_L \rangle$ designate the logical states of $\mathcal{C}$, $\mathcal{C}$ protects against perturbations by the $B_{\alpha}$ if and only if \cite[Eq. 6, 7]{knill:97}:
\begin{equation} \label{qecc-conditions}
\begin{gathered}
\langle 0_L | B^{\dagger}_{\alpha} B_{\beta} | 1_L \rangle = 0, \\
\langle 0_L | B^{\dagger}_{\alpha} B_{\beta}  | 0_L \rangle = \langle 1_L | B^{\dagger}_{\alpha} B_{\beta}  | 1_L \rangle\,.
\end{gathered}
\end{equation}
\noindent
As discussed in \cite{knill:97}, the $B_{\alpha}$ are in practice not fully known, so a practical code can only be proved protective against some anticipated subset of the $B_{\alpha}$.

The conditions given in Eq. \eqref{B-op-def} and \eqref{qecc-conditions} are completely generic, requiring only that the $B_{\alpha}$ do not act on $|S \rangle$ prior to its encoding by $E$ \cite{knill:97}; see also \cite{kribs:05} for further discussion, and in particular \cite{laflamme:96} which exhibits a means of perfecting QECCs.

Hence we can represent any QECC compactly as a mapping:
\begin{equation}
S \xrightarrow{E} \mathcal{C} \xrightarrow{U_{\mathcal{C}}} \mathcal{C} \xrightarrow{D} S^{\prime}
\end{equation}
\noindent
in which $U_{\mathcal{C}}$ is an automorphism of $\mathcal{C}$ (technically, of $\mathcal{H}_{\mathcal{C}}$), $S^{\prime}$ is the sector onto which the encoded data are decoded, and the composition $D U_{\mathcal{C}} E$ is required to be information-preserving.  We can, therefore, represent a QECC generically as a TQFT that implements the quantum channel in a LOCC protocol:
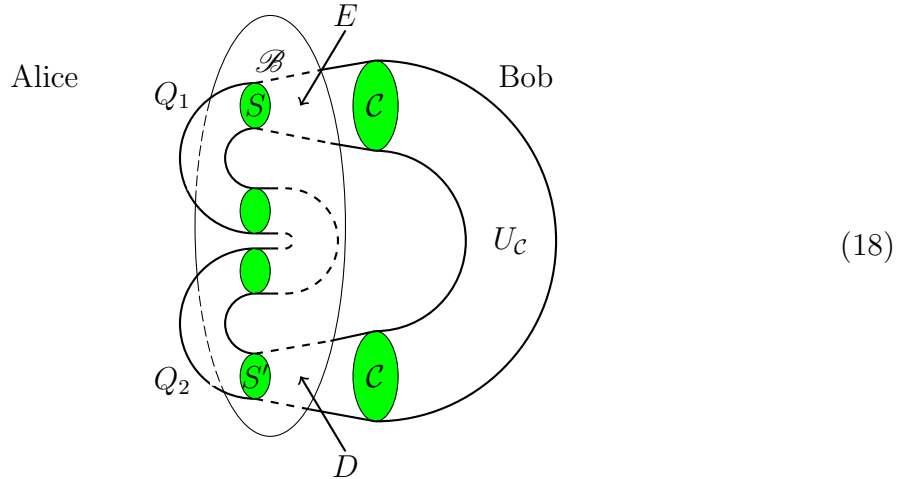
\begin{equation} \label{locc-qecc}
\begin{gathered}
\begin{tikzpicture}[every tqft/.append style={transform shape}]
\draw[rotate=90] (0,0) ellipse (2.8cm and 1 cm);
\node[above] at (0,1.9) {$\mathscr{B}$};
\draw [thick] (-0.2,1.9) arc [radius=1, start angle=90, end angle= 270];
\draw [thick] (-0.2,1.3) arc [radius=0.4, start angle=90, end angle= 270];
\draw[rotate=90,fill=green,fill opacity=1] (1.6,0.2) ellipse (0.3 cm and 0.2 cm);
\draw[rotate=90,fill=green,fill opacity=1] (0.2,0.2) ellipse (0.3 cm and 0.2 cm);
\draw [thick] (-0.2,-0.3) arc [radius=1, start angle=90, end angle= 270];
\draw [thick] (-0.2,-0.9) arc [radius=0.4, start angle=90, end angle= 270];
\draw[rotate=90,fill=green,fill opacity=1] (-0.6,0.2) ellipse (0.3 cm and 0.2 cm);
\draw[rotate=90,fill=green,fill opacity=1] (-2.0,0.2) ellipse (0.3 cm and 0.2 cm);
\draw [rotate=180, thick, dashed] (-0.2,0.9) arc [radius=0.7, start angle=90, end angle= 270];
\draw [rotate=180, thick, dashed] (-0.2,0.3) arc [radius=0.1, start angle=90, end angle= 270];
\draw [thick] (-0.2,0.5) -- (0,0.5);
\draw [thick] (-0.2,-0.1) -- (0,-0.1);
\draw [thick] (-0.2,-0.9) -- (0,-0.9);
\draw [thick] (-0.2,-0.3) -- (0,-0.3);
\draw [thick, dashed] (0,0.5) -- (0.2,0.5);
\draw [thick, dashed] (0,-0.1) -- (0.2,-0.1);
\draw [thick, dashed] (0,-0.9) -- (0.2,-0.9);
\draw [thick, dashed] (0,-0.3) -- (0.2,-0.3);
\node[above] at (-3,1.7) {Alice};
\node[above] at (3.4,1.7) {Bob};
\draw [ultra thick, white] (-0.9,1.5) -- (-0.7,1.5);
\draw [ultra thick, white] (-1,1.3) -- (-0.8,1.3);
\draw [ultra thick, white] (-1,1.1) -- (-0.8,1.1);
\draw [ultra thick, white] (-1,0.9) -- (-0.8,0.9);
\draw [ultra thick, white] (-1.1,0.7) -- (-0.8,0.7);
\draw [ultra thick, white] (-1.1,0.5) -- (-0.8,0.5);
\draw [ultra thick, white] (-1,-0.9) -- (-0.8,-0.9);
\draw [ultra thick, white] (-1,-1.1) -- (-0.8,-1.1);
\draw [ultra thick, white] (-1,-1.3) -- (-0.8,-1.3);
\draw [ultra thick, white] (-0.9,-1.5) -- (-0.7,-1.5);
\draw [ultra thick, white] (-0.9,-1.7) -- (-0.7,-1.7);
\draw [ultra thick, white] (-0.8,-1.9) -- (-0.6,-1.9);
\draw [ultra thick, white] (-0.8,-2.1) -- (-0.6,-2.1);
\node[above] at (-1.3,1.4) {$Q_1$};
\node[above] at (-1.3,-2.4) {$Q_2$};
\draw[rotate=90,fill=green,fill opacity=1] (1.6,-1.4) ellipse (0.6 cm and 0.3 cm);
\draw[rotate=90,fill=green,fill opacity=1] (-2.0,-1.4) ellipse (0.6 cm and 0.3 cm);
\draw [rotate=180, thick] (-1.4,2.6) arc [radius=2.4, start angle=90, end angle= 270];
\draw [rotate=180, thick] (-1.4,1.4) arc [radius=1.2, start angle=90, end angle= 270];
\draw [thick, dashed] (-0.2,1.9) -- (0.7,2.08);
\draw [thick] (0.7,2.08) -- (1.4,2.2);
\draw [thick, dashed] (-0.2,1.3) -- (0.8,1.1);
\draw [thick] (0.8,1.1) -- (1.4,1);
\draw [thick, dashed] (-0.2,-1.7) -- (0.8,-1.53);
\draw [thick] (0.8,-1.53) -- (1.4,-1.4);
\draw [thick, dashed] (-0.2,-2.3) -- (0.52,-2.44);
\draw [thick] (0.52,-2.44) -- (1.4,-2.6);
\node at (-0.2,1.6) {$S$};
\node at (-0.2,-2) {$S^{\prime}$};
\node at (1.4,1.6) {$\mathcal{C}$};
\node at (1.4,-2) {$\mathcal{C}$};
\node at (3.2,-0.2) {$U_{\mathcal{C}}$};
\node at (1,2.8) {$E$};
\draw [thick, ->] (1,2.6) -- (0.4,1.6);
\node at (1,-3.2) {$D$};
\draw [thick, ->] (1,-3) -- (0.4,-2);
\end{tikzpicture}
\end{gathered}
\end{equation}
\noindent

Some general remarks are now in order:

{\bf Remark 1}: Any QECC requires a classical channel between its users if it is to have practical utility.  It is the existence of this classical channel, in particular, that operationally demonstrates the separability of the users.  Hence in practice, QECCs are both required by, and require, LOCC protocols.

{\bf Remark 2}: While the above description treats the encoding operator $E$ and its inverse $D$ as implemented by Bob, in practice it is more common to think of these operators as QRFs implemented by Alice.  In this case, the codespace $\mathcal{H}_{\mathcal{C}}$ becomes a sector of the boundary $\mathscr{B}^{\prime}$ between Alice$^{\prime}$ -- i.e. Alice plus the degrees of freedom needed to implement $E$ and $D$ -- and Bob$^{\prime}$.  ``Moving the boundary'' from $\mathscr{B}$ to $\mathscr{B}^{\prime}$ is simply moving the ``Heisenberg cut'' between the observer and the observed system; it changes nothing about the physics.  Hence whether we regard the observers or ``users'' of $\mathcal{C}$ as interacting directly with the codespace is a mere notational issue.  

{\bf Remark 3}: Consonant with the above, the entanglement that serves as a resource for error correction is always provided by one or both of the two interacting bulk systems, i.e. by either Alice or Bob.  The operators $E$, $U_{\mathcal{C}}$, and $D$ can all be regarded as QRFs implemented by either Alice or Bob.

{\bf Remark 4}: The ``users'' of a QECC (here, $A_1$ and $A_2$) must be separable both from each other and from whatever bulk system implements the QECC (here, Bob).  As discussed in \S \ref{separability} above, this is equivalent to noting that classical communication (or classical memory) is a required resource for a useful QECC.

\begin{example}{\textbf{Quantum Darwinism depends on a QECC}}

{\rm As discussed in \S \ref{qDar} above, quantum Darwinism requires that the composite systems $E_{X_{i}} X E_{X_{j}}$ between distinct observers $A_i$ and $A_j$ of some quantum system $X$ implement a quantum channel that enables $A_i$ and $A_j$ to each, independently, report an observation of the same state $|X \rangle$ of $X$.  Let us examine this one step at a time.  First, we note that the environmental fragments $E_{X_i}$ each provide a continuous path from $X$ to the appropriate $A_i$ in the representation of \cite{zurek:03, zurek:05}; hence we can assume that each fragment is simply connected.  The choices of the $E_{X_i}$ are arbitrary, and in particular, not optimized in any way by the $A_i$.  Redundant encoding of $|X \rangle$ by the $E_{X_i}$ then requires, as noted earlier, that the local interactions $H_{XE_{X_i}} = H_{XE_{X_j}}$ for all components $i, j$.  Hence the interaction $H_{XE_{X}}$ between $X$ and its total environment $E_X$ must be uniform, at least at the resolution sampled by the $E_{X_i}$; it cannot, in particular, depend on any local coordinates that characterize either $X$ or $E_X$.  }

{\rm Let us now consider the particular fragments $E_{X_{i}}$ and $E_{X_{j}}$, $i \neq j$, and the interactions $H_{XE_{X_i}}$ and $H_{XE_{X_j}}$ defined at the boundaries $\mathscr{B}_i$ and $\mathscr{B}_j$ between $E_{X_{i}}$ and $E_{X_{j}}$, respectively, and $X$.  The requirement that $H_{XE_{X}}$ be uniform is the requirement that $E_{X_{i}}$ and $E_{X_{j}}$ employ the same basis for their local components of the overall interaction $H_{XE_{X}}$; otherwise the eigenvalues of $H_{XE_{X}}$ would depend on the fragment chosen.  The fragments $E_{X_{i}}$ and $E_{X_{j}}$ are, therefore, entangled \cite{fgm:22a}.  It is this entanglement that enforces $H_{XE_{X_i}} = H_{XE_{X_j}}$ and enables -- indeed, enforces -- redundant encoding of $|X \rangle$ by $E_{X_{i}}$ and $E_{X_{j}}$. }

{\rm Any state $|X \rangle$ encoded by any of the $E_{X_{i}}$ must be an eigenstate of the relevant interaction $H_{XE_{X_i}}$ and hence a component of an eigenstate of $H_{XE_{X}}$; this requirement is termed {\em einselection} in \cite{zurek:03}.  Consider now the consequences of small perturbations of $|X \rangle$ that perturb $H_{XE_{X}}$ and hence the encoded eigenvalues.  We can represent such perturbations as the actions of error-inducing operators $B_\alpha$ that transform, for example, an eigenstate of position into a spatial superposition \cite{zurek:06, zurek:09}.  Such perturbations are damped out, on average, simply by their stochasticity; in the terminology of \cite{zurek:03, zurek:05}, this stochastic damping is referred to as the {\em prediction sieve} implemented by the $E_{X_i}$.  The prediction sieve effectively implements the trace ${\rm Tr}_E (\rho_{XE_{X}})$ over environmental degrees of freedom found in standard environmental-decoherence theory \cite{zurek:03}.  The encoding is described at equilibrium by a TQFT, while the role of the stochastic perturbations can be included as an out-of-equilibrium QFT-described phase, which finally undergoes relaxation toward the TQFT phase by following a geometric relaxation flow \cite{Lulli:2021bme,Lulli:2023fcl}. }

{\rm We can, therefore, see quantum Darwinism as the observation that the environment $E_X$ of any quantum system $X$ is, effectively, a codespace that redundantly encodes eigenvalues of $H_{XE_{X}}$ and damps out the actions of any ``noise'' operators $B_\alpha$ that perturb those eigenvalues by perturbing $|X \rangle$.  This picture can be effectively extended, so as to account for error-inducing stochastic perturbations acting on the system $X$, exploiting out-of-equilibrium renormalization group flow dynamics of complex systems  \cite{Lulli:2021bme,Lulli:2023fcl}.  The codespace is ``large'' in the sense of allowing multiple observers to access the encoded state information without perturbing the encoding.  The encoding is uniform, and hence redundant, because $H_{XE_{X}}$ is uniform.  Entanglement between fragments $E_{X_i}$ of $E_X$ -- which is undetectable by any observer(s) restricted to one or a few of the $E_{X_i}$ -- assures uniformity of $H_{XE_{X}}$ and hence is the key resource required for redundant encoding. }

\end{example}

\begin{example}{\textbf{Bell/EPR experiments depend on a QECC}}

{\rm Bell/EPR experiments are some of the most important in contemporary physics \cite{aspect:81, georgescu:21}.  Using the notation above, the system $X$ in a Bell/EPR experiment is an entangled pair, with the state $|X \rangle$ being the singlet $(|\uparrow \downarrow \rangle - |\downarrow \uparrow \rangle)/\surd 2$.  There are only two observers, $A_1$ and $A_2$.  Each observer $A_i$ interacts with a fragment of the environment $E_{X_i}$ comprising a detector, etc.  The observers exchange results via a classical channel that enables classical statistical tests, e.g. tests for the violation of Bell's inequality. }

{\rm A Bell/EPR experiment can, therefore, be seen as an instance of quantum Darwinism that involves two observers interacting with a singlet state.  What enables the detection of entanglement, and hence gives the experiment its power, is the arrangement of the environmental fragments $E_{X_1}$ and $E_{X_2}$ in a way that allows each of $A_1$ and $A_2$ to access only one component of the singlet; this is accomplished, in practice, by spatially separating the detectors.  This restriction of the observers to single state components allows the actions of one observer to influence the outcome observed by other.  It effectively converts the QECC described in the example above into an erasure code.  To see this, let us examine the experiment in stages.  Provided the $E_{X_i}$ are ``set'' to employ the same $z$ axis, the combined environment $E_X$ damps out the actions of mutually-uncorrelated perturbations $B_\alpha$ of $|X \rangle$ as discussed above.  The detectors, in this case, ``witness'' the arrival of ``particles'' with classically correlated spins.  Allowing the observers to each freely choose a $z$ axis, and hence a measurement basis, between observations introduces a distinct set of perturbations $B^{\prime}_\alpha$.  While operationally, each of the $B^{\prime}_\alpha$ perturbs only the local environmental fragment $E_{X_i}$, Eq. \eqref{qecc-conditions} requires each of the $B^{\prime}_\alpha$ to act on the entire state $|X \rangle$.  It is this requirement that encodes the effective entanglement\footnote{Note that Entanglement Purification Protocols on certain mixed states sharing halves of EPR pairs through a channel, can also yield a QECC, and conversely \cite{bennett:96}.} between $E_{X_1}$ and $E_{X_2}$, and hence that a state component ``erased'' by measurement at $E_{X_1}$ is reproduced, with a spin flip, at $E_{X_2}$. }
\end{example}

\section{Conclusion}

We have shown here that TQFTs provide a minimal-assumption language for describing quantum-state preparation and measurement, and hence a minimal-assumption language for describing the multi-agent interactions that implement LOCC protocols.  We then showed how LOCC protocols generically induce QECCs, and considered how QECCs are induced in multi-observer settings described by quantum Darwinism and in two-observer Bell/EPR experiments.  These examples both illustrate the dependence of a QECC on a uniform choice of basis, and hence the dependence of a QECC on LOCC.  We can, therefore, see LOCC protocols and QECCs as generically equivalent in practice: a LOCC protocol is useless unless the quantum channel is robust against noise and hence functions as a QECC, while a QECC is useless to observers who have no means of agreeing what basis to employ.  

It is interesting to note that in both quantum Darwinism and Bell/EPR experiments, the bulk entanglement induces, via the action of a QECC, redundant classical encodings at spatially-separated locations, i.e. at the locations of spatially-separated observers.  As pointed out by Einstein \cite{einstein:48}, spatial separation is a resource for separability, and hence for effective classicality.  The generality of the methods employed in \S \ref{qecc} suggests that this outcome is completely general: that bulk entanglement generically induces classical redundancy via spatial separation.  The task of showing that this is the case -- that spacetimes can, generically, be regarded as QECCs -- is taken up in the accompanying Part II of this paper.  This result yields an intriguing hypothesis, one consistent with remarks already made in \cite{addazi:21}: that the fundamental role of spacetime in physics is to provide ``a place to put'' redundancy.  Indeed it suggests that the concept of redundancy as a necessary resource for communication and the concept of spacetime as a necessary resource for dynamics may, at some fundamental level, be the same concept.  This being the case, it would substantially support Wheeler's contention \cite{wheeler:83} that physics is fundamentally about information exchange; a striking assertion that the methods of this paper point towards with significance. The overall train of ideas further suggests support for the suggestion of Grinbaum \cite{grinbaum:17} that physics is also fundamentally about language.

%%%%%%%%%%%%%%%%%%%%%%%%%%%%%%%%%%%%%%%%%%%%%%%%%%%%%%%%%%%%%%%%%%%%%%%%%%%%%%%%%%%%%%%%%%%%%%%%%%%%%%%%%%

\section*{Appendix A: The Basics of Channel Theory Information Flow}\label{channel-app}

\subsection*{A.1: Classifiers and Infomorphisms}\label{channel-app-1}

{\em Channel Theory}, as the foundational architecture of information flow in \cite{barwise:97}, commences with the idea of a {\em classifier} (or {\em classification}) as specifying a {\em context} in terms of its constituent {\em tokens} in some language and the {\em types} to which they belong.
\begin{definition}\label{class-def-1}
A {\em classifier} $\mathcal{A}$ is a triple $\langle Tok(\mathcal{A}), Typ(\mathcal{A}), \models_{\mathcal{A}} \rangle$ where $Tok(\mathcal{A})$ is a set of {\em tokens}, $Typ(\mathcal{A})$ is a set of {\em types}, and $\models_{\mathcal{A}}$ is a {\em classification} relation between tokens and types.
\end{definition}
This definition specifies a (Barwise-Seligman) classifier/classification as an object in the category of {\em Chu spaces} originally conceived by Barr and his student Chu in \cite{barr:79},
where `$\models_{\mathcal{A}}$' is realized by a satisfaction relation between objects and attributes that is valued in some, {\it a priori}, structure-less set $\mathbf{K}$ (see also \cite{pratt:99a}). The corresponding Chu space object is simply $\A = (\rm{object}, r, \rm{attribute})$, where $r: \rm{object} \times \rm{attribute} \lra \mathbf{K}$ is the Chu space valuation of the classification (satisfaction relation).

An {\em infomorphism} is a pair of maps $\overrightarrow{f}$ and $\overleftarrow{f}$ between classifiers $\mathcal{A}$ and $\mathcal{B}$ such that the following diagram commutes:
\begin{equation}\label{info-diagram-1}
\xymatrix@!C=3pc{\rm{Typ}(\A) \ar[r]^{\overrightarrow{f}}   & \rm{Typ}(\B) \ar@{-}[d]^{\Vdash_{\B}} \\
\rm{Tok}(\A) \ar@{-}[u]^{\Vdash_{\A}}  & \rm{Tok}(\B) \ar[l]_{\overleftarrow{f}}}
\end{equation}
Definition \eqref{info-diagram-1} above represents conforming  to a set of logical constraints between classifiers. Spelt out intuitively,
an infomorphism ``transmits the information'' from one classifier to another, so that, e.g. ``b is type B'' can encode or represent the information ``a is type A''. Infomorphisms are further amenable as such when the local logics of a (regular) theory are taken into account, leading to {\em logic infomorphisms} \cite[\S 12]{barwise:97} (reviewed in \cite{fg:19a,fg:22}); as an example of a (regular) theory:
\
\medn
\textbf{Example}
\medn
{\em A first order language $L$} is a classifier, where ${Tok}(L)$ consists of a set $M$ of certain mathematical structures, and
${Typ}(L)$ are sentences in $L$, and $M \Vdash \varphi$, if and only if $\varphi$ is true in the token $M$. The type set of a token $M$ is the set of all sentences of $L$ true in $M$, called the \emph{theory} of $M$ \cite[Ex 2.2, p.28]{barwise:97}.

The Channel Theory of \cite{barwise:97} specifies information as inherently a physical mode of distinctions and relationships between them, and not simply series of bits as characterizing the passive signaling of Shannon information in neglecting a substance of reasoning. In light of these distinctions,
\cite{collier:11} applies the logical formulism of Channel Theory to argue that causation itself may be viewed as a form of computation in view of the regular relations in a distributed system. More specifically, for any pair of classifiers $\A,\B$ in a CCCD, there exists a (logic) infomorphism between them such that $\A$ directly `causes' $\B$ in terms of their respective Tokens and Types, and classifications $\Vdash_{\A}, \Vdash_{\B}$.
With classifiers as objects and
infomorphisms as arrows, Channel Theory constitutes a category isomorphic to Chu spaces as objects and Chu morphisms as arrows, respectively. Substantial applications of
Channel Theory/Chu spaces to theoretical computer science, higher dimensional automata, and e.g. conceptual/event space modeling can be found in \cite{pratt:99a,pratt:00,pratt:97} (see also \cite{fg:19a,fg:19b,fg:22} that include surveying the theories with respect to formal concept analysis, ontologies, and contextuality, respectively).

\subsection*{A.2: Algebraic operations on classifiers}\label{operations-1}

There are several significant algebraic operations on Chu spaces that are directly applicable to Barwise-Seligman classifiers \cite{pratt:97,pratt:00} (and references therein) when replacing the Chu space satisfaction relations (valuations) $r,s$ by $\Vdash_{\A}, \Vdash_{\B}$, respectively, etc.
\medn
\textbf{1.}~ Consider two Chu spaces $\A = (A,r,X)$ and $\B = (B, s, Y)$. The asynchronous parallel composition or \emph{concurrence} $\A \Vert \B$ is defined as $(A + B, t, X \times Y)$ where $t(a,(x,y)) = r(a,x)$ and $t(b,(x,y)) = s(b,y)$. The operational sense of this relation is the process that
$\A$ and $\B$ occur independently of each other.
\medn
\textbf{2.}~The tensor product or \emph{orthocurrence} $\A \otimes \B$ of two processes  $\A = (A,r,X)$ and $\B = (B, s, Y)$, is defined as $(A \times B, t, Z)$ where $Z$ denotes the set of Chu transforms from $\A$ to the transpose $\B^{\perp} = (Y, \hat{r}, B)$ of $\B$ ($\hat{r}(y,b) = r(b,y))$, and
$t((a,b), (f,g)) = s(b, f(a))$. The sense is that of ``flow through'', e.g. as a system $\A$ of trains passes through a system $\B$ of stations, or a system of signals flowing through a gated circuit, or two particle systems in collision. As for $\B \otimes \A$, this follows from the appropriate replacement
in the definition, and up to bijection $\A \otimes \B$ and $\B \otimes \A$ are isomorphic as Chu spaces (i.e. they have the same carrier, cocarrier and matrix) \cite{pratt:97,pratt:00}.
\medn
\textbf{3.}~The \emph{choice} $\A \sqcup \B$ of two processes $\A = (A,r,X)$ and $\B = (B, s, Y)$ is defined as $(A + B, t, X + Y)$ where $t(a,x) = r(a,x), ~t(b,y) = s(b,y)$, and $t(a,y) = t(b,x) =0$. This describes how the events of $\A \sqcup \B$  are formed as the disjoint union of those of $\A$ and $\B$, prescribing partitioning into two sorts: those assigning $0$ to all events of $\B$ choosing to execute $\A$, any state allowable in $\A \sqcup \B$, and conversely, choosing $\B$ and then setting all events of $\A$ equal to $0$.

\subsection*{A.3: Cocone Diagram (CCD)}\label{channel-app-2}

When compared with the isomorphic category of Chu spaces and Chu morphisms, Channel Theory follows a separate, but intimately tied direction: infomorphisms as mappings between classifiers provides the basic building blocks towards constructing multi-level, quasi-hierarchical, distributional classification systems described in \cite{barwise:97}. Typically, one starts with a commuting
{\em cocone diagram} (CCD) of infomorphisms between classifiers:
\begin{equation}\label{ccd-1}
\xymatrix@C=4pc{&\mathbf{C^\prime} &  \\
\A_1 \ar[ur]^{f_1} \ar[r]_{g_{12}} & \A_2 \ar[u]_{f_2} \ar[r]_{g_{23}} & \ldots ~\A_k \ar[ul]_{f_k}
}
\end{equation}
where $\mathbf{C^\prime}$ denotes the {\em core}, or the {\em colimit} when defined \cite{awodey:10}.
The commutativity requirement of \eqref{ccd-1} makes explicit the role of $\mathbf{C^\prime}$ as a ``wire'' or shared memory for the component classifiers \cite{barwise:97,allwein:04}.

\subsection*{A.4: Cone-Cocone Diagram (CCCD)}\label{cccd}

The diagram of a commuting finite {\em cone of infomorphisms} (CD) represents the dual construction, in which all the arrows are reversed.  In this case the core of the (dual) channel is the limit of all possible downward-going structure-preserving maps to the classifiers $\A_i$. Combining a (commuting) CCD with a (commuting) CD leads to a {\em cone-cocone diagram} (CCCD), i.e.

\begin{equation}\label{cccd-1}
\xymatrix@C=4pc{&\mathbf{C^\prime} &  \\
\A_1 \ar[ur]^{f_1} \ar[r]_{g_{12}}^{g_{21}} & \ar[l] \A_2 \ar[u]_{f_2} \ar[r]_{g_{23}}^{g_{32}} & \ar[l] \ldots ~\A_k \ar[ul]_{f_k} \\
&\mathbf{D^\prime} \ar[ul]^{h_1} \ar[u]^{h_2} \ar[ur]_{h_k}&
}
\end{equation}
which informationally serves as a hierarchical {\em memory-write system} (see e.g. \cite{fg:19a,fg:19b,fg:22}), such as seen
in the blue-print for a typical variational auto-encoder (VAE):
\begin{equation}\label{cccd-2}
\begin{gathered}
\xymatrix@C=6pc{\mathcal{A}_1 \ar[r]_{g_{12}}^{g_{21}} & \ar[l] \mathcal{A}_2 \ar[r]_{g_{23}}^{g_{32}} & \ar[l] \ldots ~\mathcal{A}_k \\
&\mathbf{C^\prime} \ar[ul]^{h_1} \ar[u]^{h_2} \ar[ur]_{h_k}& \\
\mathcal{A}_1 \ar[ur]^{f_1} \ar[r]_{g_{12}}^{g_{21}} & \ar[l] \mathcal{A}_2 \ar[u]_{f_2} \ar[r]_{g_{23}}^{g_{32}} & \ar[l] \ldots ~\mathcal{A}_k \ar[ul]_{f_k}
}
\end{gathered}
\end{equation}

\subsection*{A.5: Shared QRFs: CCCDs under concurrency}\label{concurrency}

Let us consider two CCCDs inducing two shared QRFs: these will be $\rm{CCCD}_{\A}$ inducing $\rm{QRF}_{\A}$,
and $\rm{CCCD}_{\B}$ inducing $\rm{QRF}_{\B}$. Next, we take the operation of concurrency `$\Vert$' in \S\ref{operations-1},
defining $\Ce_k = \A_i \Vert \B_j$ (for admissible $i,j,k \geq 0$). This is illustrated below for the respective CCD sections with their corresponding infomorphisms (here denoted the same for convenience):

\bign
\textbf{CCD for} $\mathbf{QRF_{\A}}$ :
\bign
\begin{equation}\label{concurrent-1}
\xymatrix@C=4pc{&\mathbf{C^\prime}_{\A} &  \\
\A_1 \ar[ur]^{f_1} \ar[r]_{g_{12}} & \A_2 \ar[u]_{f_2} \ar[r]_{g_{23}} & \ldots ~\A_k \ar[ul]_{f_k}
}
\end{equation}

\bign
\textbf{CCD for} $\mathbf{QRF_{\B}}$ :
\bign
\begin{equation}\label{concurrent-2}
\xymatrix@C=4pc{&\mathbf{C^\prime}_{\B} &  \\
\B_1 \ar[ur]^{f_1} \ar[r]_{g_{12}} & \B_2 \ar[u]_{f_2} \ar[r]_{g_{23}} & \ldots ~\B_k \ar[ul]_{f_k}
}
\end{equation}

\bign
$\mathbf{CCD~for~entangled~QRF_{\Ce} = QRF_{\A \Vert \B}}$ :
\bign
\begin{equation}\label{concurrent-3}
\xymatrix@C=4pc{&\mathbf{C^\prime}_{\Ce} &  \\
\Ce_1 \ar[ur]^{f_1} \ar[r]_{g_{12}} & \Ce_2 \ar[u]_{f_2} \ar[r]_{g_{23}} & \ldots ~\Ce_k \ar[ul]_{f_k}
}
\end{equation}

%%%%%%%%%%%%%%%%%%%%%%%%%%%%%%%%%%%%%%%%%%%%%%%%%%%%%%%%%%%%%%%%%%%%%%%%%%%%%%%%%%%%%%%%%%%%%%%%%%%%%%%%%

\section*{Appendix B: Sequential measurements and TQFT}\label{tqft}

A TQFT can be represented as a functor from the category of cobordisms to the category of Hilbert spaces \cite{atiyah:88,quinn:95}.  We prove in \cite{fgm:22} that any QRF can be represented as a CCCD, and then construct a category with CCCDs as objects and morphisms of CCCDs, which must by definition respect the commutativity of CCCDs as diagrams, as morphisms.  The category is, effectively, a category of QRFs, in which the morphisms represent sequential choices of QRF to be applied to the data encoded on some sector $S$.  We show that all such choices can be represented by one of two diagrams.  Using the compact notation

\begin{equation} \label{QRF-1}
\begin{gathered}
\begin{tikzpicture}
\draw [thick] (0,0) -- (2,1) -- (2,-1) -- (0,0);
\node at (1.3,0) {S};
\end{tikzpicture}
\end{gathered}
\end{equation}
\noindent
to represent a QRF $S$, we can represent measurements of a physical situation in which one system divides into two, possibly entangled, systems with a diagram of the form

\begin{equation} \label{QRF-2}
\begin{gathered}
\begin{tikzpicture}
\node at (0,0) {$S$};
\draw [thick] (0.2,0) -- (1,1) -- (2,1.5) -- (2,0.5) -- (1,1);
\node at (1.7,1) {$S_1$};
\draw [thick] (0.2,0) -- (1,-1) -- (2,-0.5) -- (2,-1.5) -- (1,-1);
\node at (1.7,-1) {$S_2$};
\end{tikzpicture}
\end{gathered}
\end{equation}
\noindent
Parametric down-conversion of a photon exemplifies this kind of process.  The reverse process can be added to yield:

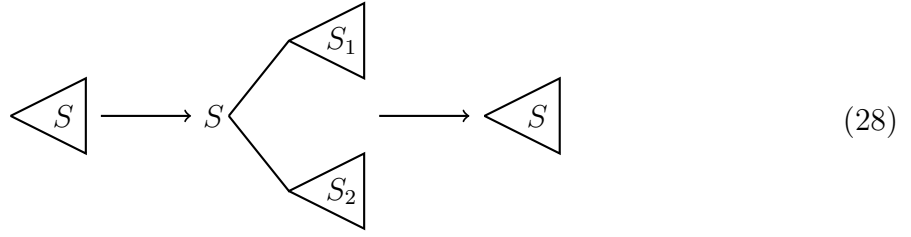
\begin{equation} \label{flow-1}
\begin{gathered}
\begin{tikzpicture}
\node at (0,0) {$S$};
\draw [thick] (0.2,0) -- (1,1) -- (2,1.5) -- (2,0.5) -- (1,1);
\node at (1.7,1) {$S_1$};
\draw [thick] (0.2,0) -- (1,-1) -- (2,-0.5) -- (2,-1.5) -- (1,-1);
\node at (1.7,-1) {$S_2$};
\draw [thick] (-2.7,0) -- (-1.7,0.5) -- (-1.7,-0.5) -- (-2.7,0);
\node at (-2,0) {$S$};
\draw [thick, ->] (-1.5,0) -- (-0.3,0);
\draw [thick, ->] (2.2,0) -- (3.4,0);
\draw [thick] (3.6,0) -- (4.6,0.5) -- (4.6,-0.5) -- (3.6,0);
\node at (4.3,0) {$S$};
\end{tikzpicture}
\end{gathered}
\end{equation}
\noindent
Diagram \eqref{flow-1} represents a relabelling of subsets of the base-level classifiers that act on the sector $S$:

\begin{equation} \label{class-flow-1}
\underbrace{\mathcal{A}_1, \mathcal{A}_2, \dots \mathcal{A}_m}_{S} \rightarrow \underbrace{\mathcal{A}_1, \dots \mathcal{A}_i,}_{S_1} \underbrace{\mathcal{A}_{i+1}, \dots \mathcal{A}_m}_{S_2} \rightarrow \underbrace{\mathcal{A}_1, \mathcal{A}_2, \dots \mathcal{A}_m}_{S}
\end{equation}
\noindent
In the second type of sequential measurement process, the pointer-state QRF $P$ is replaced with an alternative QRF $Q$ with which it does not commute.  Sequences in which position and momentum, $s_z$ and $s_x$ are measured alternately are examples.  These can be represented by the diagram

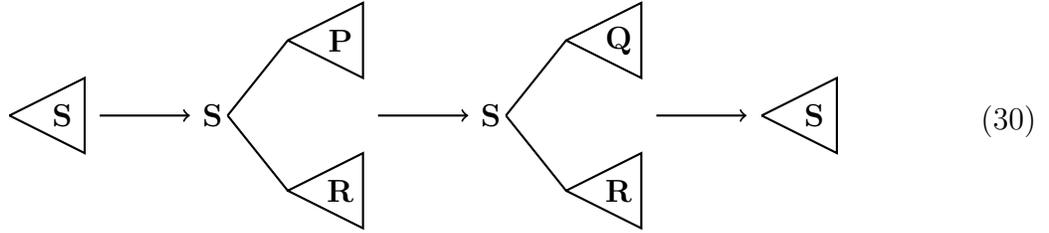
\begin{equation} \label{flow-2}
\begin{gathered}
\begin{tikzpicture}
\node at (0,0) {$\mathbf{S}$};
\draw [thick] (0.2,0) -- (1,1) -- (2,1.5) -- (2,0.5) -- (1,1);
\node at (1.7,1) {$\mathbf{P}$};
\draw [thick] (0.2,0) -- (1,-1) -- (2,-0.5) -- (2,-1.5) -- (1,-1);
\node at (1.7,-1) {$\mathbf{R}$};
\draw [thick] (-2.7,0) -- (-1.7,0.5) -- (-1.7,-0.5) -- (-2.7,0);
\node at (-2,0) {$\mathbf{S}$};
\draw [thick, ->] (-1.5,0) -- (-0.3,0);
\draw [thick, ->] (2.2,0) -- (3.4,0);
\node at (3.7,0) {$\mathbf{S}$};
\draw [thick] (3.9,0) -- (4.7,1) -- (5.7,1.5) -- (5.7,0.5) -- (4.7,1);
\node at (5.4,1) {$\mathbf{Q}$};
\draw [thick] (3.9,0) -- (4.7,-1) -- (5.7,-0.5) -- (5.7,-1.5) -- (4.7,-1);
\node at (5.4,-1) {$\mathbf{R}$};
\draw [thick, ->] (5.9,0) -- (7.1,0);
\draw [thick] (7.3,0) -- (8.3,0.5) -- (8.3,-0.5) -- (7.3,0);
\node at (8,0) {$\mathbf{S}$};
\end{tikzpicture}
\end{gathered}
\end{equation}
\noindent
Again this can be written as a relabeling of classifiers, leaving the pointer-state classifiers that are traced over when measuring only the reference component $R$ for system identification implicit, as:

\begin{equation} \label{class-flow-2}
\underbrace{\mathcal{A}_1, \mathcal{A}_2, \dots \mathcal{A}_k}_{R} \rightarrow \underbrace{\mathcal{A}_1, \dots \mathcal{A}_k,}_{R} \underbrace{\mathcal{A}_{k+1}, \dots \mathcal{A}_m}_{P} \rightarrow \underbrace{\mathcal{A}_1, \dots \mathcal{A}_k,}_{R} \underbrace{\tilde{\mathcal{A}}_{k+1}, \dots \tilde{\mathcal{A}}_m}_{Q} \rightarrow \underbrace{\mathcal{A}_1, \mathcal{A}_2, \dots \mathcal{A}_k}_{R}
\end{equation}
\noindent
where the notation $\tilde{\mathcal{A}}_{l}$ indicates that $\mathcal{A}_{l}$ has been rewritten in a rotated measurement basis, e.g. $s_z \rightarrow s_x$ or $x \rightarrow p= m\, (\partial x/ \partial t)$.  As both $P$ and $Q$ must commute with $R$, the commutativity requirements for $S$ are satisfied.

Measurement sequences of the form of Diagram \eqref{flow-1} can be mapped to cobordisms of the form:

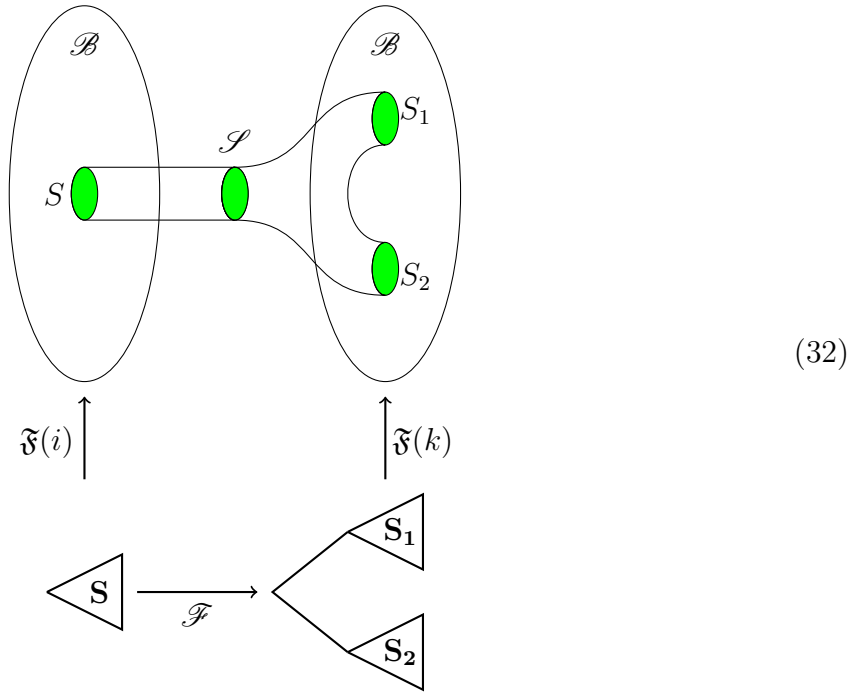
\begin{equation} \label{CCCD-to-Cob}
\begin{gathered}
\begin{tikzpicture}[every tqft/.append style={transform shape}]
\draw[rotate=90] (0,0) ellipse (2.5cm and 1 cm);
\node[above] at (0,1.7) {$\mathscr{B}$};
\node at (-0.4,0) {$S$};
\begin{scope}[tqft/every boundary component/.style={draw,fill=green,fill opacity=1}]
\begin{scope}[tqft/cobordism/.style={draw}]
\begin{scope}[rotate=90]
\pic[tqft/cylinder, name=a];
\pic[tqft/pair of pants, anchor=incoming boundary 1, name=b, at=(a-outgoing boundary 1)];
\end{scope}
\end{scope}
\end{scope}
\draw[rotate=90] (0,-4) ellipse (2.5cm and 1 cm);
\node[above] at (4,1.7) {$\mathscr{B}$};
\node at (2,0.7) {$\mathscr{S}$};
\node at (4.4,1.1) {$S_1$};
\node at (4.4,-1.1) {$S_2$};
\draw [thick, <-] (0,-2.7) -- (0,-3.8);
\draw [thick, <-] (4,-2.7) -- (4,-3.8);
\draw [thick] (-0.5,-5.3) -- (0.5,-4.8) -- (0.5,-5.8) -- (-0.5,-5.3);
\node at (0.2,-5.3) {$\mathbf{S}$};
\draw [thick] (3.5,-4.5) -- (4.5,-4) -- (4.5,-5) -- (3.5,-4.5);
\node at (4.2,-4.5) {$\mathbf{S_1}$};
\draw [thick] (3.5,-6.1) -- (4.5,-5.6) -- (4.5,-6.6) -- (3.5,-6.1);
\node at (4.2,-6.1) {$\mathbf{S_2}$};
\draw [thick] (3.5,-4.5) -- (2.5,-5.3) -- (3.5,-6.1);
\draw [thick, ->] (0.7,-5.3) -- (2.3,-5.3);
\node at (-0.5,-3.3) {$\mathfrak{F}(i)$};
\node at (4.5,-3.3) {$\mathfrak{F}(k)$};
\node at (1.5,-5.6) {$\mathscr{F}$};
\end{tikzpicture}
\end{gathered}
\end{equation}

while sequences of the form of Diagram \eqref{flow-2} can be mapped to cobordisms of the form:

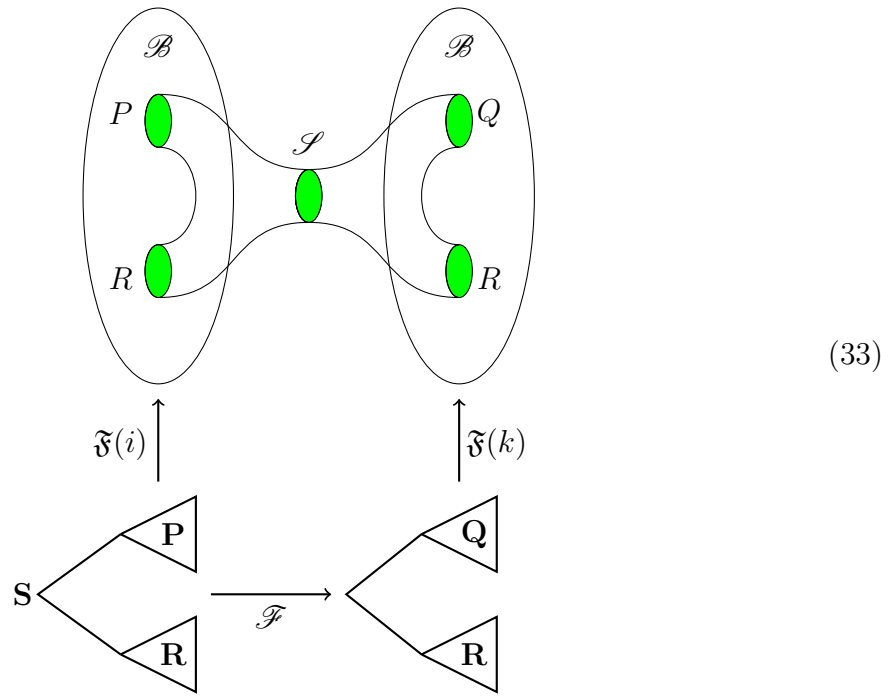
\begin{equation} \label{CCCD-to-Cob-2}
\begin{gathered}
\begin{tikzpicture}[every tqft/.append style={transform shape}]
\draw[rotate=90] (0,0) ellipse (2.5cm and 1 cm);
\node[above] at (0,1.7) {$\mathscr{B}$};
\node at (-0.5,1.1) {$P$};
\node at (-0.5,-1.1) {$R$};
\begin{scope}[tqft/every boundary component/.style={draw,fill=green,fill opacity=1}]
\begin{scope}[tqft/cobordism/.style={draw}]
\begin{scope}[rotate=90]
\pic[tqft/reverse pair of pants, at={(-1,0)}, name=a];
\pic[tqft/pair of pants, anchor=incoming boundary 1, name=b, at=(a-outgoing boundary 1)];
\end{scope}
\end{scope}
\end{scope}
\draw[rotate=90] (0,-4) ellipse (2.5cm and 1 cm);
\node[above] at (4,1.7) {$\mathscr{B}$};
\node at (2,0.7) {$\mathscr{S}$};
\node at (4.4,1.1) {$Q$};
\node at (4.4,-1.1) {$R$};
\draw [thick, <-] (0,-2.7) -- (0,-3.8);
\draw [thick, <-] (4,-2.7) -- (4,-3.8);
\draw [thick] (-0.5,-4.5) -- (0.5,-4) -- (0.5,-5) -- (-0.5,-4.5);
\node at (0.2,-4.5) {$\mathbf{P}$};
\draw [thick] (-0.5,-6.1) -- (0.5,-5.6) -- (0.5,-6.6) -- (-0.5,-6.1);
\node at (0.2,-6.1) {$\mathbf{R}$};
\draw [thick] (-0.5,-4.5) -- (-1.6,-5.3) -- (-0.5,-6.1);
\node at (-1.8,-5.3) {$\mathbf{S}$};
\draw [thick] (3.5,-4.5) -- (4.5,-4) -- (4.5,-5) -- (3.5,-4.5);
\node at (4.2,-4.5) {$\mathbf{Q}$};
\draw [thick] (3.5,-6.1) -- (4.5,-5.6) -- (4.5,-6.6) -- (3.5,-6.1);
\node at (4.2,-6.1) {$\mathbf{R}$};
\draw [thick] (3.5,-4.5) -- (2.5,-5.3) -- (3.5,-6.1);
\draw [thick, ->] (0.7,-5.3) -- (2.3,-5.3);
\node at (-0.5,-3.3) {$\mathfrak{F}(i)$};
\node at (4.5,-3.3) {$\mathfrak{F}(k)$};
\node at (1.5,-5.6) {$\mathscr{F}$};
\end{tikzpicture}
\end{gathered}
\end{equation}

In either case, $\mathfrak{F}: \mathbf{CCCD} \rightarrow \mathbf{Cob}$ is the required functor from the category $\mathbf{CCCD}$ of CCCDs to the category of $\mathbf{Cob}$ finite cobordisms.  In general, we can state:

\begin{theorem}[\cite{fgm:22} Thm. 1] \label{thm2}
For any morphism $\mathscr{F}$ of CCCDs in $\mathbf{CCCD}$, there is a cobordism $\mathscr{S}$ such that a diagram of the form of Diagram \eqref{CCCD-to-Cob} or \eqref{CCCD-to-Cob-2} commutes.
\end{theorem}
\noindent
referring to \cite{fgm:22} for the proof.

%%%%%%%%%%%%%%%%%%%%%%%%%%%%%%%%%%%%%%%%%%%%%%%%%%%%%%%%%%%%%%%%%%%%%%%%%%%%%%%%%%%%%%%%%%%%%%%%%%%%%%%%%

\section*{Appendix C: Rice's theorem}

Rice's theorem \cite{rice:53} states that only trivial properties of computing programs are algorithmically decidable, i.e. decidable in finite time by a Turing machine with arbitrarily-large memory.

We begin by distinguishing the class {\bf RE} of recursively-enumerable languages from the smaller class {\bf R} of recursive languages.  A language $\mathscr{L}$ is in {\bf RE} if a Turing machine can decide in finite time if a given finite string $s$ is in $\mathscr{L}$.  A language $\mathscr{L}$ is in {\bf R} if a Turing machine can decide in finite time if a given finite string $s$ is or is not in $\mathscr{L}$.  See \cite{hopcroft:79} for background and details.

The Halting Problem is the problem of deciding, for any program (language) $\mathscr{L}$ and any input string $s$ whether $s$ is in $\mathscr{L}$, i.e. whether the program $\mathscr{L}$ ``halts'' by outputting either `1' (or ``accept'' indicating $s \in \mathscr{L}$) or `0' (or ``reject'' indicating $s \notin \mathscr{L}$).  This problem is undecidable \cite{turing:37}; hence {\bf RE} $\supset$ {\bf R}.  Again see \cite{hopcroft:79} for background and details.

We can state Rice's theorem informally as:

\begin{theorem*}[Rice \cite{rice:53} (informal)] \label{rice-informal}
Whether an arbitrarily-chosen program $P$ implements a nontrivial function $y = f(x)$ is (finite Turing) undecidable.
\end{theorem*}
\noindent
and prove it by showing that it reduces to the Halting problem:

\begin{proof}
(Sketch) If one can decide that an arbitrary $P$ computes $y = f(x)$, one can decide whether the program $P^{\prime}$ specified by:

\begin{align*}
& P \leftarrow x \\
& \mathrm{If} ~(y = f(x)), ~\mathrm{Halt}
\end{align*}
\noindent
halts.  But whether an arbitrary $P^{\prime}$ halts is undecidable \cite{turing:37}.
\end{proof}

More formally, we can define a ``property'' $P$ as a set of languages, i.e. say that any language $\mathscr{L}$ that satisfies the property $P$, belongs to $P$, namely $\mathscr{L} \in P$. A property $P$ is {\em trivial} if: i) $P$ is not satisfied by any $\mathscr{L} \in$ {\bf RE}; or ii) $P$ is satisfied by all $\mathscr{L} \in$ {\bf RE}. Conversely, a property is {\em non-trivial} if it is satisfied by some $\mathscr{L}_i \in$ {\bf RE}, and is not satisfied by some $\mathscr{L}_j \in$ {\bf RE}, $i \neq j$.  Therefore, for a non-trivial property $P$, there exist Turing machines $M_i$ and $M_j$ such that $M_i$ has $P$ but $M_j$ does not.

With these definitions, we can state the theorem formally:

\begin{theorem*}[Rice \cite{rice:53}] \label{rice-formal}
If $P$ is a non-trivial property of an {\bf RE} language, then:
\begin{equation}
\mathscr{L}_{P} = \{\langle M \rangle | \mathscr{L}(M) \in P \}
\end{equation}
is undecidable.
\end{theorem*}

The following proof is adapted from \cite{hopcroft:79}:

\begin{proof}
Consider a property $P$ such that the empty language $\emptyset \notin P$; the proof for the case of $\emptyset \in P$ is analogous.  Since $P$ is nontrivial, $\exists \mathscr{L} \neq \emptyset, \mathscr{L} \in P$.  Let $M_\mathscr{L}$ be a Turing machine that accepts $\mathscr{L}$, i.e. accepts all strings in $\mathscr{L}$.

Let $M$ be an arbitrary Turing machine, let $w$ be an arbitrary string, and let $x \in \mathscr{L}$.  We now construct a Turing machine $M^\prime$ that performs the following operations:

\begin{itemize}
\item $M^\prime$ simulates $M$ acting on the input $w$;
\item if the simulation of $M$ acting on $w$ halts on `reject', $M^\prime$ halts;
\item if the simulation of $M$ acting on $w$ halts on `accept', $M^\prime$ reads $x$ as input;
\item if $M^\prime$ reads $x$ as input, $M^\prime$ simulates $M_\mathscr{L}$ on $x$.
\end{itemize}
\noindent
We can now observe:
\begin{itemize}
\item if $M$ rejects $w$, $M^\prime$ never reads $x$; hence the language accepted by $M^\prime$ is $\emptyset$.  Since $\emptyset \notin P$, $M^\prime$ does not have $P$.
\item if $M$ accepts $w$, $M^\prime$ accepts $x$.  By definition, $x$ is an arbitrary string in $\mathscr{L}$; hence $M^\prime$ accepts $\mathscr{L}$.  By definition, $\mathscr{L} \in P$, so $M^\prime$ does have $P$.
\end{itemize}
\noindent
But $M$ and $w$ are arbitrary, so $M^\prime$ both has and does not have $P$.  Contradiction.
\end{proof}

\section*{Acknowledgments}
A.M.~wishes to acknowledge support by the Shanghai Municipality, through the grant No.~KBH1512299, by Fudan University, through the grant No.~JJH1512105, the Natural Science Foundation of China, through the grant No.~11875113, and by the Department of Physics at Fudan University, through the grant No.~IDH1512092/001.

\end{document}